\documentclass[twoside,twocolumn,9pt]{article}
\usepackage{extsizes}
\usepackage[super,sort&compress,comma]{natbib}
\usepackage[version=3]{mhchem}
\usepackage[left=1.5cm, right=1.5cm, top=1.785cm, bottom=2.0cm]{geometry}
\usepackage{balance}
\usepackage{mathptmx}
\usepackage{sectsty}
\usepackage{graphicx}
\usepackage{lastpage}
\usepackage[format=plain,justification=justified,singlelinecheck=false,font={stretch=1.125,small,sf},labelfont=bf,labelsep=space]{caption}
\usepackage{float}
\usepackage{fancyhdr}
\usepackage{fnpos}
\usepackage[english]{babel}
\addto{\captionsenglish}{%
  
}
\usepackage{array}
\usepackage{droidsans}
\usepackage{charter}
\usepackage[T1]{fontenc}
\usepackage[usenames,dvipsnames]{xcolor}
\usepackage{setspace}
\usepackage[compact]{titlesec}
\usepackage{hyperref}

\usepackage{dcolumn}

\usepackage[dvipsnames]{xcolor}
\usepackage{chemformula}
\usepackage[normalem]{ulem}
\usepackage{txfonts}

\newcommand{\cm}{cm$^{-1}$}
\newcommand{\icm}{cm\textsuperscript{-1}}
\newcommand{\ai}{\textit{ab initio}}
\newcommand{\xa}{$\tilde{X}^{1}A_{1}$}
\newcommand{\ab}{$\tilde{A}^{1}B_{1}$}

\newcommand{\trove}{\textsc{TROVE}}

\newcommand{\sihh}{SiH$_{2}$}
\newcommand{\silane}{SiH$_{4}$}
\newcommand{\disil}{Si$_{2}$H$_{6}$}

\newcommand{\forward}{SiH$_{2}$~+~SiH$_{4}$ $\rightarrow$ Si$_{2}$H$_{6}$}
\newcommand{\reverse}{Si$_{2}$H$_{6}$ $\rightarrow$ SiH$_{2}$~+~SiH$_{4}$}

\newcommand{\xc}{ExoCross}

\newcommand{\TROVE}{{\textsc{TROVE}}}

\newcommand{\Cv}[1]{${\mathcal C}_{#1{\rm v}}$}
\newcommand{\Cs}{${\mathcal C}_{\rm s}$}

\newcommand{\Dd}[1]{${\mathcal D}_{#1{\rm d}}$}

\newcommand{\ket}[1]{\vert #1 \rangle  }

\DeclareFontFamily{U}{matha}{\hyphenchar\font45}

\newcommand{\2}{$_{2}$}
\newcommand{\3}{$_{3}$}
\newcommand{\4}{$_{4}$}
\newcommand{\6}{$_{6}$}

\newcolumntype{d}[1]{D{.}{.}{#1}}

\definecolor{cream}{RGB}{222,217,201}

\graphicspath{{./Figures/}}

\begin{document}

\pagestyle{fancy}
\thispagestyle{plain}
\fancypagestyle{plain}{
\renewcommand{\headrulewidth}{0pt}
}

\makeFNbottom
\makeatletter
\renewcommand\LARGE{\@setfontsize\LARGE{15pt}{17}}
\renewcommand\Large{\@setfontsize\Large{12pt}{14}}
\renewcommand\large{\@setfontsize\large{10pt}{12}}
\renewcommand\footnotesize{\@setfontsize\footnotesize{7pt}{10}}
\makeatother

\renewcommand{\thefootnote}{\fnsymbol{footnote}}
\renewcommand\footnoterule{\vspace*{1pt}%
\color{cream}\hrule width 3.5in height 0.4pt \color{black}\vspace*{5pt}}
\setcounter{secnumdepth}{5}

\makeatletter
\renewcommand\@biblabel[1]{#1}
\renewcommand\@makefntext[1]%
{\noindent\makebox[0pt][r]{\@thefnmark\,}#1}
\makeatother
\renewcommand{\figurename}{\small{Fig.}~}
\sectionfont{\sffamily\Large}
\subsectionfont{\normalsize}
\subsubsectionfont{\bf}
\setstretch{1.125} 
\setlength{\skip\footins}{0.8cm}
\setlength{\footnotesep}{0.25cm}
\setlength{\jot}{10pt}
\titlespacing*{\section}{0pt}{4pt}{4pt}
\titlespacing*{\subsection}{0pt}{15pt}{1pt}

\fancyfoot{}
\fancyfoot[LO,RE]{\vspace{-7.1pt}\includegraphics[height=9pt]{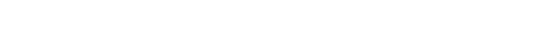}}
\fancyfoot[CO]{\vspace{-7.1pt}\hspace{11.9cm}\includegraphics{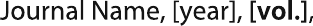}}
\fancyfoot[CE]{\vspace{-7.2pt}\hspace{-13.2cm}\includegraphics{head_foot/RF}}
\fancyfoot[RO]{\footnotesize{\sffamily{1--\pageref{LastPage} ~\textbar  \hspace{2pt}\thepage}}}
\fancyfoot[LE]{\footnotesize{\sffamily{\thepage~\textbar\hspace{4.65cm} 1--\pageref{LastPage}}}}
\fancyhead{}
\renewcommand{\headrulewidth}{0pt}
\renewcommand{\footrulewidth}{0pt}
\setlength{\arrayrulewidth}{1pt}
\setlength{\columnsep}{6.5mm}
\setlength\bibsep{1pt}

\makeatletter
\newlength{\figrulesep}
\setlength{\figrulesep}{0.5\textfloatsep}

\newcommand{\topfigrule}{\vspace*{-1pt}%
\noindent{\color{cream}\rule[-\figrulesep]{\columnwidth}{1.5pt}} }

\newcommand{\botfigrule}{\vspace*{-2pt}%
\noindent{\color{cream}\rule[\figrulesep]{\columnwidth}{1.5pt}} }

\newcommand{\dblfigrule}{\vspace*{-1pt}%
\noindent{\color{cream}\rule[-\figrulesep]{\textwidth}{1.5pt}} }

\makeatother

\twocolumn[
  \begin{@twocolumnfalse}
{\includegraphics[height=30pt]{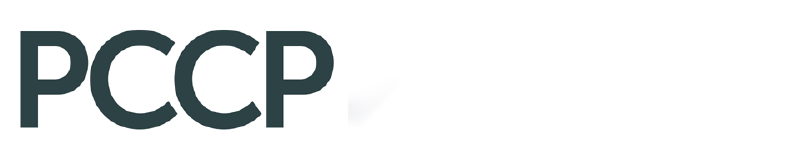}\hfill\raisebox{0pt}[0pt][0pt]{\includegraphics[height=55pt]{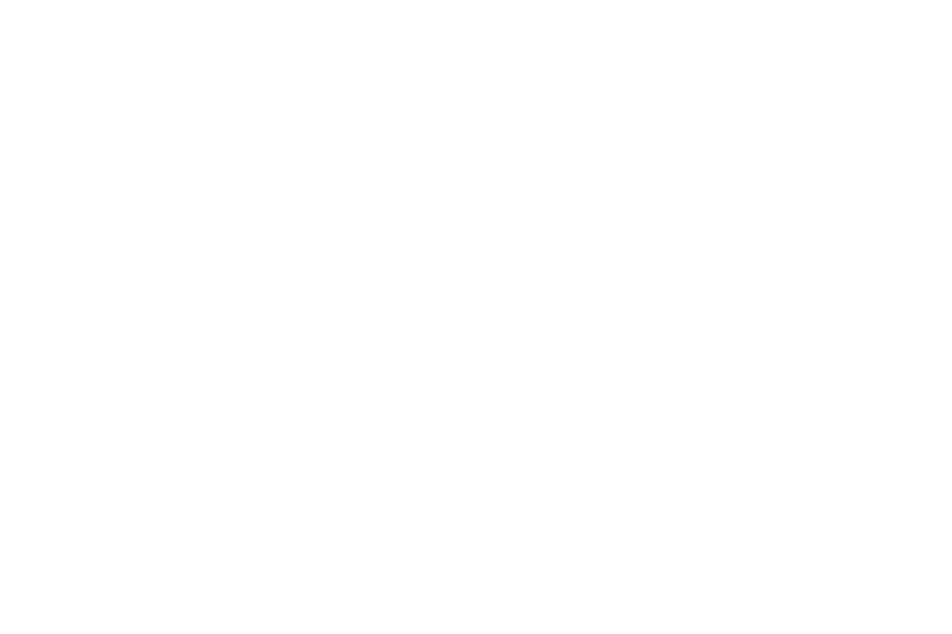}}\\[1ex]
\includegraphics[width=18.5cm]{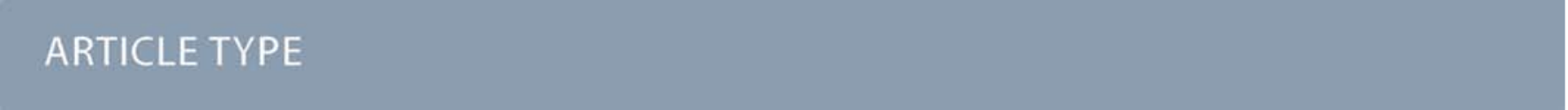}}\par
\vspace{1em}
\sffamily
\begin{tabular}{m{4.5cm} p{13.5cm} }

\includegraphics{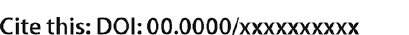} & \noindent\LARGE{\textbf{Modelling  the non-local thermodynamic equilibrium  spectra of silylene (SiH$_{2}$)$^\dag$}} \\
\vspace{0.3cm} & \vspace{0.3cm} \\

 & \noindent\large{Victoria H.J. Clark\textit{$^{\ast}$} and Sergei N. Yurchenko$^{\ast\ast}$ } \\

\includegraphics{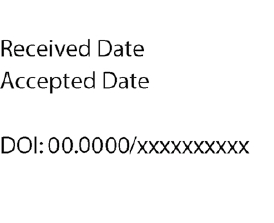} & \noindent\normalsize{
This paper sets out a robust methodology for modelling  spectra of polyatomic molecules produced in reactive or dissociative environments, with vibrational populations outside local thermal equilibrium ({LTE}). The methodology is based on accurate, extensive ro-vibrational line lists containing transitions with high vibrational excitations and relies on the detailed ro-vibrational assignments. The developed methodology is applied to model non-LTE IR and visible spectra of silylene (SiH$_2$) produced  in  a  decomposition of disilane (Si$_2$H$_6$), a reaction of technological importance. Two approaches  for non-LTE vibrational populations of the product SiH$_2$ are introduced: a simplistic 1D approach based on the Harmonic approximation and a full 3D model incorporating accurate  vibrational wavefunctions of SiH$_2$  computed variationally with the  TROVE (Theoretical ROVibrational Energy) program. We show how their non-LTE spectral signatures can be used to trace different reaction  channels of molecular dissociations.  } \\

\end{tabular}

 \end{@twocolumnfalse} \vspace{0.6cm}

  ]

\renewcommand*\rmdefault{bch}\normalfont\upshape
\rmfamily
\section*{}
\vspace{-1cm}


\footnotetext{\textit{Department of Physics and Astronomy, University College London, Gower Street, WC1E 6BT London, United Kingdom}}


\footnotetext{$^{\ast}$~Corresponding author; E-mail: v.clark.17@ucl.ac.uk}
\footnotetext{$^{\ast\ast}$E-mail: s.yurchenko@ucl.ac.uk}
\footnotetext{\dag~Electronic Supplementary Information (ESI) available. See DOI: 10.1039/b000000x/}



\date{\today}



\section{Introduction}
\label{sec:intro}


Normally, molecules are assumed to be in local thermal equilibrium pertaining to a given temperature with the internal degrees of freedom (electronic-rotation-vibration) characterized by the Boltzmann distribution. However many different physical chemical, experimental and technological processes produce molecules that do not satisfy the Boltzmann law and as consequence have unusual, non-local thermal equilibrium (non-LTE) spectroscopic signatures.
Molecules produced in reactions do not necessarily obey the Boltzmann thermal equilibrium, at least if the reaction time is shorter than the collision time.  Instead, their internal degrees of freedom are populated based on the reaction paths rather than on the  temperature of the surrounding environment.
These out-of-LTE (i.e. non-LTE) populations encode  information about the structural reaction dynamics and  can manifest in the molecular spectra. The field of non-LTE spectroscopy has great potential to  study these processes as the  properties of the molecules producing the non-LTE spectroscopic signatures can shed the light on the dynamics of chemical reactions.\citep{82HeLexx.CH3,11FeKuKa.HCN, 14FeMiSh.HCN, 17FeKuKn.HCN, 20PrBaCh.HCN,21PaCiCl.CO} The so-called transition state (TS) spectroscopy is a technique already widely used that employs the high-resolution non-LTE spectra of products to observe reaction processes that  are hidden for the conventional  spectroscopic methods. \citep{96WeHrBo,92GrMoPo.CO}
The novel high resolution non-LTE spectroscopic techniques allow  decoupling of the vibrational and rotational degrees of freedom of molecules  and thus control their vibrational and rotational populations, e.g. with rich vibrational and simplified rotational structures.\citep{01MiKeAn.H2O,01MiKeWa.H2O,01CoSpBu.NO,14ZhDoLi,20DuSuBr.CH4,21PaCiCl.CO}

The modern day study of non-LTE spectroscopy can be traced as far back as the 1930s to the original papers of \citet{30Milnex}, and the many key papers from the decades following.\citep{56CuGoxx, 69Hought, 69KuLoxx, 72Dickin, 74KuJaxx, 75Shvedx, 86LoRoLo, 86LoRoMo, 92WiPiSh, williams.thesis}  The non-LTE spectroscopic effects play important role in high-resolution applications and there exist a number of accurate  non-LTE spectroscopic and radiative transfer codes, see \citet{07vaBlSc,GRANADA,19PaLaxx} and references therein.  As such, non-LTE spectra are often vital for the modelling of astrophysical problems, including planetary atmospheric properties\citep{01LoTaxx}, stellar atmospheres of solar system and exoplanets\citep{05Asplund,60Chandr,64Feautr,71Rybick,72Athay,78Mihala,93SiCrxx,21ChZhYo} and the ISM. \citep{99GoLaxx.ISM,07vaBlSc,09LivaKl.NO}

The non-local thermodynamic effects of the spectra of molecules has been of interest to chemists and astronomers alike for many years. A notable example of this is the 2011 work by \citet{11FeKuKa.HCN} who studied the isomers of HCN within acetonitrile, formamide, and BrCN discharge.
Using the features for both the HCN and HNC molecules from these spectra, \citet{11FeKuKa.HCN} were able to calculate the ratios of molecules within the reactions and also the reaction path taken by the HNC molecule during the isomerization. In this work we explore this idea to study non-LTE spectral signatures of  silylene (SiH$_2$) produced from disilane (Si$_2$H$_6$).

Reaction properties of silylene, silane and disilane such as the rate constants for the formation, destruction and chemical pathways are important for plasma physics aspects such as silicon deposition.\citep{88JaChxx.SiH2,90DoDoGa.SiH2,87JaMeSc.SiH2} The ease of hydrogen transfer and high barriers in the saturated silicon system, leading to the ready formation of three-centre interactions and consequently the isomerisation reactions of \disil, are just as important to study as the elementary reactions.\citep{93ErSaKa.SiH2}  Silane  containing reactions are also of importance for astrophysics, with the presence of \silane\ in IRC~+10216 discussed by \citet{84GoBexx.SiH4,05KaOsxx.SiH4}.

\begin{figure}[ht]
\centering
\includegraphics[width=\columnwidth]{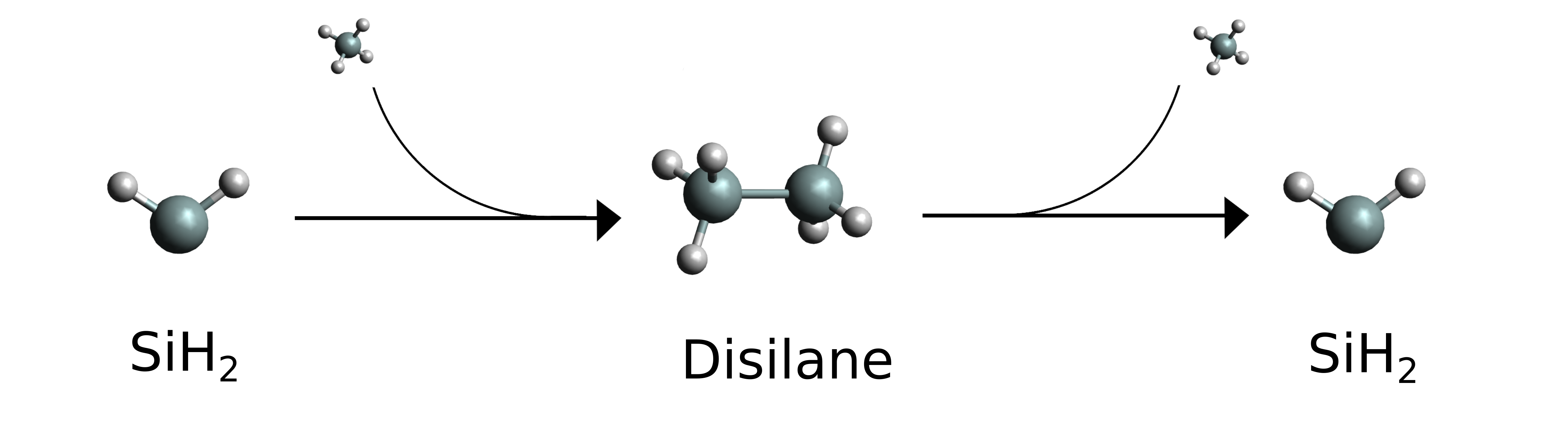}
\caption{Schematic reaction path for the silylene.  Molecular \sihh\ is the starting material, fragment \sihh\ is the product. }
\label{fig:path-scheme}
\end{figure}

The complexity of the silane containing systems has been discussed elsewhere \citep{88AgThRa.SiH2, 93ErSaKa.SiH2}, with quantitative calculations proving particularly difficult. There were a number of \ai\ studies of the structural properties of  \disil\ $^{\textrm{e.g.,}}$ \citep{95BeFrMa.SiH2,05PuTaxx.Si2H6,94ChDuMo.Si2H6,95Sunxxx.Si2H6,91MaSaFe.Si2H6,02RoKiRa.Si2H6,97ChRiPa,90AgThRa.Si2H6} as well as of the formation and reactions involving this molecule.\citep{03HuWaWa, 12LaFrPe.SiH2, 12EgGeRo.Si2H6, 06YoMaOg, 02AnWaxx.Si2H6} \citet{90AgThRa.Si2H6}  produced  a global  \ai\ potential energy surface of disilane and used it to investigate the dissociation dynamics with  classical trajectories. \citet{91MaSaFe.Si2H6}  reported a force field for Si\2H\6.  The main isomer of \disil\ has a staggered, ethane-like,  structure (see Fig.~\ref{fig:path-scheme})  with a low barrier (1.2 kcal/mol) to the eclipsed conformation  \citep{80DuChxx}. \disil\ has been shown to have a local minimum as an inverted stable structure with one of the Si--H$_3$ `umbrellas' pointing to the  center as well as a transition state with a similar inverted configuration.\citep{93SaNaxx,95BeFrMa.SiH2} These structures are nominally asymmetric (\Cs\ symmetry) but essentially acquiring the \Cv{3}\ symmetry.

Thermal decomposition of \disil\ has been extensively studied, both theoretically (mostly using RRKM, Rice-Ramsperger-Kassel-Marcus) and experimentally, with the reaction \reverse\ as the most common \citep{12EgGeRo.Si2H6,06YoMaOg,90AgThRa.Si2H6,71BoPuxx.Si2H6} and important decomposition process of the excited disilane \citep{84OlPoRe,90MaONRi,92MoJeCa,95MiMaRo,06YoMaOg,01OnPaxx.SiH4,90AgThRa.Si2H6}, and where  Arrhenius parameters and rate constants have been reported  (e.g. \citet{87RoJeCa,90MaONRi,95MiMaRo,71BoPuxx.Si2H6}).
However, it is also possible for the disilane molecule to dissociate homolytically, as was originally thought to be the main pathway owing to disilanes similarities with ethane, and form 2\ch{SiH3^{.}},\citep{12EgGeRo.Si2H6,90AgThRa.Si2H6,39EmRexx.Si2H6} or to undergo dehydrogenation to H$_{2}$Si-SiH$_{2}$, H$_{3}$Si-SiH, or  H$_{3}$Si-H$_{2}$.\citep{90AgThRa.Si2H6,03HuWaWa,06YoMaOg} Disilane can even undergo double dehydrogenation to form Si$_{2}$H$_{2}$, however to our knowledge this has only been reported as the main product when undergoing photolysis at 193~nm.\citep{99TaToMa.Si2H6} \citet{06YoMaOg} notes that the transition state for the \reverse\ is 8.48~kcal~mol$^{-1}$ lower than the transition state for Si$_{2}$H$_{6}$ $\rightarrow$ H$_{3}$SiSiH~+~H$_{2}$, at 43.38~kcal~mol$^{-1}$ compared to 51.86~kcal~mol$^{-1}$.

The spectroscopy of SiH\2\ has been used to monitor the \forward\  and  \reverse\ reactions (see Fig.~\ref{fig:path-scheme}) and measure the corresponding rate constants and Arrhenius parameters by spectroscopically tracking electronic ($\tilde{A}$-$X$) transitions of SiH\2\ \citep{88JaChxx.SiH2,94MaRoxx.SiH2,95MiMaRo,95BeFrMa.SiH2}. In these studies, the reconstructions of the amount of \sihh\ relied on the assumption of the Boltzmann thermal distribution when estimating the population of the lower state.
No account of the possible non-LTE population of  \sihh\  molecules after dissociation  was  made, which could potentially hamper the count of the \sihh\ molecules and effect the reaction rates estimated.
A similar experimental technique  was used in \citet{00HeJoxx.SiH2} to monitor \sihh\ in SiH\4\ plasma.

It is the second, dissociation, part of the reaction  shown in Fig.~\ref{fig:path-scheme} (\reverse) we  study in this work. More specifically, we  show that (i) the (vibrational) populations of the molecules produced in  reactions can be very different from the Boltzmann distribution and is important to take into account when interpreting spectroscopic measurements. That (ii) spectral shapes of the dissociated \sihh\ can bear  strong non-LTE character, very different from the LTE spectrum of an LTE \sihh\ sample making  it possible to distinguish between different reaction stages and even between different dissociation channels the silylene molecules it is produced from. In this work  non-LTE spectra of \sihh\ under conditions similar to dissociation processes expected in these experiments are modelled.

Recently we have computed an accurate ro-vibrational line list for \sihh, named CATS.\citep{jt779} It covers a large range of rotational and vibrational excitation, capable of modelling very hot spectra of this molecule (up to $T=$ 2000 K) as part of the ExoMol database.\citep{jt810} The CATS line list was produced using the  program \trove,\citep{TROVE,16YuYaOv.methods} which solves the nuclear motion Schr\"{o}dinger equation  variationally. The ro-vibrational energies and corresponding wavefunctions were computed using an accurate, empirically refined  potential energy surface (PES) of silane and a high-level \ai\ dipole moment surface (DMS). The ro-vibrational probabilities (in the form of Einstein A coefficients) were computed using a high level \ai\ dipole moment surface.

The study by \citet{jt779} forms the basis for the present work, where we utilize the CATS line list, wavefunctions, purpose-built numerical basis set, and the CATS computational \trove\ setup  to model non-LTE spectroscopic properties of SiH\2\ produced from dissociation of Si\2H$_6$ through different reaction channels. Using a simplified 1D Harmonic oscillator wavefunctions (see \citet{21PaCiCl.CO}) and  more sophisticated 3D vibrational CATS wavefunctions from accurate variational calculations, the non-LTE ro-vibrational  populations of SiH\2\ are generated and used to produce non-LTE spectroscopic spectra of different dissociation channels of disilane.
To this end we investigate reaction topology  connecting the global minimum of Si\2H$_6$ with the closest saddle points and local minima as well as the corresponding structural properties using a high level \ai\ theory  cc-pVTZ-F12/CCSD(T)-f12b\citep{07AdKnWe.ai,09KnAdWe} employing the program MOLPRO2015.~\citep{MOLPRO}

Theoretically, the non-LTE properties of dissociating molecules were  studied by \citet{75BaFrxx}. In the present work we use general approach of \citet{74Berryx} and \citet{75BaFrxx},  which assume no significant structural changes between the reactant and product nuclear configuration, along with the slow vibrational relaxation of the product \citep{20DuSuBr.CH4} to investigate non-LTE spectroscopic signatures of \sihh\ produced from dissociation of Si$_2$H$_6$. In this paper we specifically consider situations where the vibrational relaxations are not achieved during the time of the experiment, so that the molecules still hold  the memory of the structure during the reaction or dissociation. The rotation relaxation time however  is much shorter and  the rotational degrees of freedom can be usually assumed to satisfy the Bolzmann equilibrium.\citep{20DuSuBr.CH4}

We also investigate possible non-LTE impact on the electronic $\tilde{A}$ (0,2,0)--$\tilde{X}$ (0,0,0) spectrum of SiH\2. This is a favorite spectroscopic system for the detection of SiH\2\ due to the large Franck-Condon factor and the availability of suitable laser.\citep{84InSuxx.SiH2,84JaWhBj.SiH2,85InSuxx.SiH2,86RaStHa.SiH2,86ThStxx.SiH2,86ObAtxx.SiH2,87ThStMc.SiH2,88JaChxx.SiH2,92FuMaOb.SiH2,93KoKoOk.SiH2,95NoAkKo.SiH2,94MaRoxx.SiH2,98EsCaxx.SiH2}

The non-LTE absorption spectra of SiH\2\ are simulated using the (non-LTE) \xc\ program, \citep{ExoCross} where a new feature of non-Boltzmann populations was added. \xc\ has been previously used to model spectra of molecules in environments that can be characterized using two temperatures, vibrational and rotational.\citep{18YuSzPy,19LiYuTe,19AdJeYa,19SmSoYu}

The paper is structured as follows. In Sec.~\ref{s:ab-initio} we describe the calculations of potential energy surfaces for the disilane and silylene structures.
The theory used in this paper is described in Sec.~\ref{sec:non-LTE}. In Sec.~\ref{sec:1d-approach} we calculate the 1D harmonic wavefunction population and use them to produce  non-LTE spectra of SiH\2\ corresponding to different dissociation routes.  In Sec.~\ref{sec:3d-approach} we calculate the populations and subsequent non-LTE using the full 3D wavefunctions and describe the new \trove\ methodology. A non-LTE electronic $\tilde{A}$ (0,2,0)--$\tilde{X}$ (0,0,0) spectrum of SiH\2 is presented in Sec.~\ref{sec:A-X-spectrum}.  Conclusions are offered in Sec.~\ref{sec:conc}.

\section{Geometry optimisation and reaction topology of Si$_2$H$_6$}
\label{s:ab-initio}

\subsection{Disilane isomers}

In order to better understand the reaction process of breaking \disil,  the topology of \disil\ has been investigated by performing a structural analysis of Si\2H\6\ using a high level  \ai\ theory. This includes finding the global minimum (GM),  local minima (LM), transition states (TS),  reaction barriers as well as reaction paths, as detailed below.  A reaction slice through the global PES of Si\2H\6\  helps to indicate how likely local minima or transitions states were to be formed based on corresponding topology.
These properties of \disil\ were obtained using the geometry optimization and reaction path finder implemented in MOLPRO2015~\citep{MOLPRO} using the  explicitly correlated coupled cluster method CCSD(T)-F12b \citep{07AdKnWe.ai,09KnAdWe} with the F12-optimized correlation consistent basis set, VTZ-F12 \citep{08PeAdWe.ai} in the frozen core approximation.  The calculations employed the diagonal fixed amplitude ansatz 3C(FIX)~\citep{04TenNo.ai} and a Slater geminal exponent value of $\beta$ = 1.0 $a_{0}^{-1}$.\citep{09HiPeKn.ai} The auxiliary basis sets were chosen to be the resolution of the identity OptRI~\citep{08YoPexx.ai} basis and the aug-cc-pV5Z/JKFIT  \citep{02Weigend.ai} and cc-awCV5Z/MP2FIT \cite{05Hattig.ai}  basis sets for density fitting.  In the following this level of theory will be referenced to as VTZ/CCSD(T)-F12b.

We shall refer to different  disilane isomers as  \textit{d}GM, \textit{d}LM and \textit{d}TS  to distinguish them from the \sihh\ fragments GM, LM and TS  as discussed below.

The global minimum  of Si\2H\6\ (\textit{d}GM) has a symmetrical, staggered \Dd{3} structure. The closest local minimum (\textit{d}LM) has an inverted, \Cv{3} structure. The lowest transition state, which will be referred to as \textit{d}TS (TS1 in \citet{95BeFrMa.SiH2} and  TS2 in \citet{02ToMuKo.SiH2}),  has also inverted structure, just a few kJ/mol above \textit{d}LM.\citep{95BeFrMa.SiH2} These structures together with the corresponding  optimized parameters for three geometries most relevant to our work are collected in Table~\ref{t:structures}. Our structural parameters compare well with that from the literature. The structure of the \textit{d}GM has also been determined  spectroscopically, with the equilibrium bond lengths  $r_{\rm Si-H}$ = 1.492~\AA, $r_{\rm Si-Si}$ = 2.331~\AA\ and bond angles $\beta_{\angle{\rm HSiSi}}$ = 110.3$^{\circ}$ and  $\alpha_{\angle{\rm HSiSi}}$ =  108.6$^{\circ}$.\citep{72BeCoFr.Si2H6} The structure of the deuterated isotopologue  Si$_{2}$H$_{5}$D was reported as  $r_{\rm Si-H}$ = 1.4874(17)~\AA, $r_{\rm Si-Si}$ = 2.3317(15)~\AA, $\beta_{\angle{\rm HSiSi}}$ =  110.66(16)$^{\circ}$.\citep{86DuHaMc.Si2H6}

\begin{figure*}[ht]
\centering
\includegraphics[width=\textwidth]{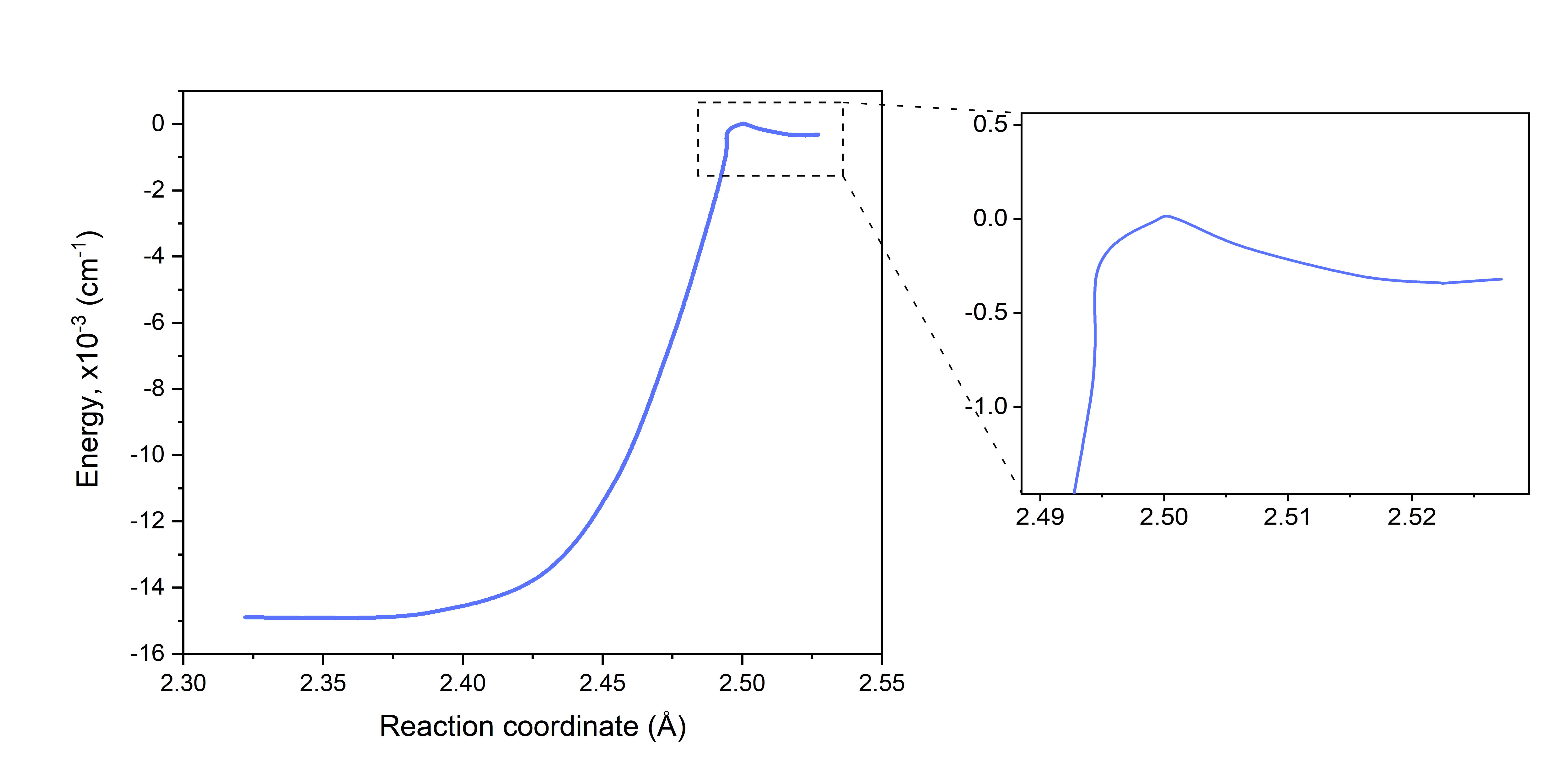}
\caption{Potential energy curve for the reaction path from disilane global minimum (\textit{d}GM), left, through the transition state (\textit{d}TS) onto the local minimum (\textit{d}LM), right.  A zoom of  \textit{d}LM to \textit{d}TS path is inset.
The \textit{d}LM is 344.22
\icm\ lower than the energy of the \textit{d}TS.
}
\label{fig:18-to-19-reaction-path-again}
\end{figure*}

The reaction path connecting the disilane isomers \textit{d}GM, \textit{d}LM and \textit{d}TS is shown in Fig.~\ref{fig:18-to-19-reaction-path-again}.  A zoom of the \textit{d}LM side is shown as inset. The energies of the global and local  minima are 178~kJ/mol and 4.1~kJ/mol below the transition state, respectively. The energy and geometry information for  \textit{d}LM, \textit{d}GM, \textit{d}TS are collected in Table \ref{t:energies}.  The results calculated compare well with both the results of \citet{95BeFrMa.SiH2} and \citet{93SaNaxx}, albeit both vary for the \textit{d}GM structure by 36 kJ~mol$^{-1}$

The \textit{d}LM isomer of \disil\ has a shallow potential with a very low barrier to \textit{d}TS of 344~\icm. It can be also recovered using lower levels of theory, for example, using MP2/6-311G(d, p) \citep{95BeFrMa.SiH2} and even with  the UFF force fields implemented in Avogadro 1.2.0 \citep{Avogadro2012} via the steepest descent method and 4 steps per update.

\newcolumntype{d}[1]{D{.}{.}{#1}}
\begin{table*}[h!]
\caption{\label{t:structures} An overview of the \disil\ molecules discussed in the paper. The subscript $L$ denotes atoms to the LHS of the Si--Si bond in the figures shown, and the subscript $R$ denotes atoms to the RHS
}
\centering
\begin{tabular}{rd{4.3}d{4.3}d{4.3}d{3.3}d{3.3}d{3.3}}
\hline
\textbf{Isomer} &  \textbf{LM} & \textbf{TS} & \textbf{GM} & \multicolumn{3}{c}{ \textbf{\citet{95BeFrMa.SiH2}}} \\ \hline
\centering{Molecule}  & \multicolumn{1}{c}{\raisebox{-\totalheight}{\includegraphics[width=0.07\textwidth]{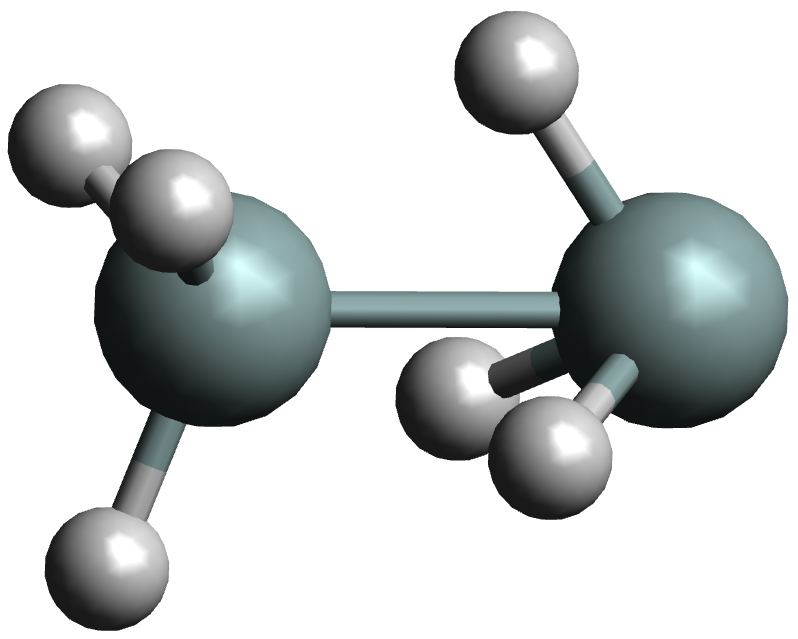}}} &
\multicolumn{1}{c}{\raisebox{-\totalheight}{\includegraphics[width=0.07\textwidth]{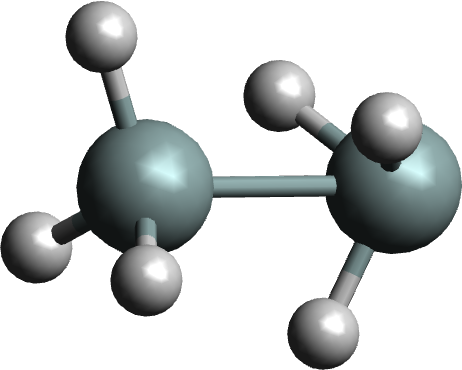}} }&
\multicolumn{1}{c}{\raisebox{-\totalheight}{\includegraphics[width=0.07\textwidth]{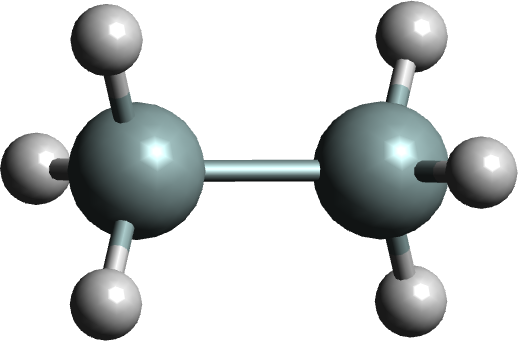}} }& \multicolumn{1}{c}{LM1} & \multicolumn{1}{c}{TS1}& \multicolumn{1}{c}{GM}\\
Si--Si (\AA) & 2.467 & 2.489 & 2.341                     &2.470 &2.485 &2.346    \\
Si$_{\rm R}$--H$_{\rm R}$ (\AA) & 1.532 & 1.499 & 1.482          &1.525 &1.493 &1.479    \\
Si$_{\rm L}$--H$_{\rm L}$ (\AA) & 1.480 & 1.480 & 1.482          &1.478 &1.478 &1.479    \\
$\angle$H$_{\rm R}$SiSi ($^{\circ}$) & 55.4 & 64.9   &110.2 &  \multicolumn{1}{c}{-- }   &   \multicolumn{1}{c}{--}   &110.24   \\
$\angle$H$_{\rm L}$SiSi ($^{\circ}$) & 110.7 & 110.0 &110.2 &110.7 &110.2 &110.24   \\
$\angle$H$_{\rm R}$SiH$_{\rm R}$ ($^{\circ}$)&90.9&100.6&108.6 & 91.4 &102.6 &    \multicolumn{1}{c}{--}     \\
$\angle$H$_{\rm L}$SiH$_{\rm L}$ ($^{\circ}$)&  108.3 &109.7&108.6 &108.2 &108.8 &    \multicolumn{1}{c}{--}     \\
$\omega_b$ (\icm) & 940 & 940 & 940 & &&  \\
$\omega_{s{\rm (L)}}$ (\icm) & 2241.69 & 2245.36 & 2240.10 & &&  \\
$\omega_{s{\rm (R)}}$ (\icm) & 2039.12 & 2170.66 & 2240.10 & &&  \\
\hline

\end{tabular}
\end{table*}

\begin{table}[h!]
\caption{\label{t:energies}The total and relative energies for the three disilane structure.  Literature values from \citet{95BeFrMa.SiH2}$^{*}$ and \citet{93SaNaxx}
}
\centering
\begin{tabular}{cd{4.8}d{4.7}}
\hline
& \multicolumn{1}{c}{\textbf{Energy, E$_{\rm{h}}$}}& \multicolumn{1}{c}{\textbf{Rel. Energy, kJ~mol${^{-1}}^{**}$ }} \\ \hline
& \multicolumn{2}{c}{\textit{d}GM}\\ \cline{2-3}
Calculated &-581.728134&-178.00\\
Becerra &-236.0&-214.50\\
Saki &-581.53962&-214.68715\\
&\multicolumn{2}{c}{\textit{d}TS}\\\cline{2-3}
Calculated &-581.660338&0.00\\
Becerra (TS1)$^{\dag}$ &-48.5&0.00\\
Saki (TS2)$^{\dag}$ &-581.45785&0.00\\
&\multicolumn{2}{c}{\textit{d}LM}\\\cline{2-3}
Calculated &-581.661906&-4.12\\
Becerra (LM1)$^{\ddag}$ &-51.3&-3.00\\
Saki (Compl. 2)$^{\ddag}$ &-581.45837&-1.3652601\\
\hline
\end{tabular}
{ \flushleft \footnotesize
$^{\dag}$: TS2 in \citet{93SaNaxx} is the same as TS1 in \citet{95BeFrMa.SiH2}\\
$^{\ddag}$: Complex 2 in \citet{93SaNaxx} is the same as LM1 in \citet{95BeFrMa.SiH2}\\
$^{*}$: relative to SiH$_2$+SiH$_4$, kJ~mol$^{-1}$\\
$^{**}$: Relative to \textit{d}TS \\
}
\end{table}

\subsection{The silylene fragments}

With the aim to give more quantitative information on the structural and dynamical properties  of five \sihh\ fragments from \disil, the fragments are described as follows. GM is an \sihh\ fragment from the global minimum structure (\textit{d}GM); LM-L is an \sihh\ fragment from the left hand side (LHS) of the local minimum structure (\textit{d}LM) with the Si--H\3\ umbrella group pointing  outside; LM-R is an \sihh\ fragment from the right hand side (RHS) of the \textit{d}LM structure with  Si--H\3\ pointing  inside; TS-L is an \sihh\ fragment from the LHS of the transition state structure (\textit{d}TS), Si--H\3\ umbrella group points  outside; and TS-R is an \sihh\ fragment from the RHS of the \textit{d}TS structure, Si--H\3\ umbrella group pointing  inside.  The structural parameters and structures are shown in Table \ref{t:structures} and the energies are shown in Table \ref{t:energies}.  The harmonic frequencies for the disilane molecules were computed with MOLPRO using the TVZ/CCSD(T)-F12b level of theory. The results from the frequency analysis are shown in Table \ref{t:freq-2} along with a comparison with literature (both experimental, if available, or theoretical data is shown).  The symmetry of \textit{d}GM is \Dd{3} whereas the symmetry for the \textit{d}LM and \textit{d}TS are both \Cv{3}. The columns titled ``Theory'' were calculated in this work, and the degenerate states have been removed, with an average calculated if there were any differences owing to computational errors associated with the lower symmetry used by MOLPRO.
All frequencies of \textit{d}LM are positive thus confirming that  it is a minimum with a   stable structure. The `negative' (or imaginary) harmonic frequency of \textit{d}TS is -551.6~\icm.

\definecolor{Gray}{gray}{0.95}

\begin{table*}[h]
\caption{\label{t:freq-2}The calculated and available literature vibrational frequencies in \icm\ for the three \disil\ molecules
}
\begin{tabular}{cd{4.2}cr@{}l|cd{4.2}cr|cd{4.2}}
\hline
\multicolumn{5}{c}{\textit{d}GM, \Dd{3} } & \multicolumn{4}{c}{\textit{d}TS, \Cv{3} } &  \multicolumn{2}{c}{\textit{d}LM, \Cv{3} }\\
\hline
\multicolumn{2}{c}{Calculated} & \multicolumn{3}{c}{Literature\cite{95BeFrMa.SiH2}} & \multicolumn{2}{c}{Calculated} & \multicolumn{2}{c}{Literature \cite{95BeFrMa.SiH2, 87RoJeCa}} & \multicolumn{2}{c}{Calculated}  \\ 
Mode & \multicolumn{1}{c}{Freq.} & Mode & \multicolumn{2}{c}{Freq.} & Mode & \multicolumn{1}{c}{Freq.} & Mode & Freq. & Mode & \multicolumn{1}{c}{Freq.} \\
\hline
E$_u$    & 2247.78 & E$_u$    & 2179& & E & 2252.06 & A & 2181 & E & 2244.30  \\
A$_{1g}$ & 2241.03 & A$_{1g}$ & 2163&$^{\dag}$ & A & 2238.66 & E & 2169 & A & 2239.08  \\
E$_g$    & 2239.28 & E$_u$    & 2155& & E & 2176.47 & A & 2105 & A & 2050.89 \\
A$_{1u}$ & 2232.30  & A$_{1u}$ & 2154& & A & 2161.85 & A & 2087 & E & 2027.35 \\
E$_u$    & 962.35  & E$_g$   & 941&$^{\dag}$ & E & 995.37 & A & 1585 & E & 1089.15  \\
E$_g$    & 948.37  & E$_u$   & 940& & E & 966.95 & A &  960 & E & 978.33  \\
A$_{1g}$ & 931.91  & A$_{1g}$ & 920&$^{\dag}$ & A & 902.23 & A &  949 & A & 956.33  \\
A$_{1u}$ & 859.02  & A$_{1u}$ & 844& & E & 570.16 & A &  927 & A & 667.30  \\
E$_g$    & 636.87 & E$_g$ & 628&$^{\dag}$ & A & 353.88 & A\&E$^{\ddag}$ &  925 & E & 618.92  \\
A$_{1g}$ & 438.76  & A$_{1g}$ & 432&$^{\dag}$ & E & 319.45 & E &  145 & A & 432.13  \\
E$_u$    & 372.85  & E$_u$ & 379& & A & 316.43 & A &  128 & A & 389.31  \\
A$_{1u}$ & 136.56  & A$_{1u}$ & 128& & A & -551.59 & E &   93 & E & 277.91  \\
\hline
\end{tabular}
{\flushleft \footnotesize
$^{\dag}$: Confirmed by the experimental gas phase Raman spectrum of \citet{80DuChxx}.\\
$^{\ddag}$:  Although there are three lines of 925~\icm, there are no triply degenerate states in the \Cv{3} symmetry.  \citet{87RoJeCa} assign one 925 \icm\ to $\varv_{15}$, which in the \textit{d}GM symmetry is an E state degenerate with $\varv_{16}$.  The \textit{d}TS $\varv_{16}$ can be found at 960 \icm.  The other two 925 \icm\ line were assigned to $\varv_{9}$ and $\varv_{10}$ (E).\\}
\end{table*}

\section{Modelling the non-LTE populations of SiH\2}
\label{sec:non-LTE}

We now consider the decomposition reaction \reverse\ shown in Fig.~\ref{fig:path-scheme} and model the vibrational populations of the product SiH\2\ assuming that the corresponding relaxation time to LTE is much longer than the time of the spectroscopic experiment. We aim at simulating non-LTE IR spectra of SiH\2\ using these  populations to demonstrate their impact on the spectral shape of dissociated species.

In our description of the non-LTE vibrational population of the dissociated molecule we follow
the polyatomic Franck-Condon type  approximation by \citet{75BaFrxx} and \citet{74Berryx}  based on the structural differences between reactant and product assuming no significant change in nuclear configuration of the molecule. In order to connect the product (gas phase SiH\2) to an initial structure of Si\2H\6\ through the dissociation process, we assume that the dissociation happens instantaneously, i.e. the initial configuration of the product SiH\2\ corresponds to the structural parameters (bond lengths Si--H and inter-bond angles $\angle$HSiH) of  SiH\2\ as a fragment of Si\2H\6, for which the parameters collected in Table~\ref{t:structures} are used. For example, for the dissociation from the \textit{d}GM structure, the initial configuration of the gas phase SiH\2\ is assumed to be  $r_{\rm Si-H}$ = 1.482~\AA, $ \alpha_{\angle{\rm HSiH}}$ = 108.6$^\circ$. Naturally, this is a very deformed geometry comparing to the  equilibrium structure of the gas phase SiH\2, $r_{\rm e}$  = 1.5137 $\pm 0.003$~\AA\ and $\alpha_{\rm e}$ = 92.04 $\pm 0.05$~$^\circ$.\citep{16KoMaSt.SiH2}
After being dissociated, in relaxing to be  a free molecule, the fragment \sihh\  has added vibrational energy and is hence in a situation when it's vibrational populations do not match the LTE distribution for corresponding temperature of the surroundings, at least for the vibrational degrees of freedom.

Experience shows that the rotational equillibration time is usually very short
and we can thus safely assume the LTE conditions for the rotational degrees of freedom with the rotational temperature the same as the temperature of the surroundings (see also e.g. \citet{20DuSuBr.CH4}).
The vibrational population however is not in the thermal equilibrium and therefore no sensible vibrational temperature could be associated with the corresponding population.

The non-LTE population $N_{J,k,\varv}(T)$ of a ro-vibrational state $\ket{J,k,\varv}$ is then given by:\citep{21PaCiCl.CO}
\begin{equation}
    \label{e:population}
 N_{J,k,\varv}(T) = \frac{g_{i}^{\rm ns}(2J+1) e^{-\frac{c_2 \tilde{E}_{J,k}^{\rm rot}}{T_{\rm rot}} }}{Q_{\rm nLTE}} N_{\varv}^{\rm vib} ,
\end{equation}
where  $N_{\varv}^{\rm vib}$ is a non-LTE vibrational population, $T_{\rm rot}$ is the rotational temperature, $J$ is the total angular momentum quantum number; $k$ is a generic rotational quantum number, e.g. the projection of the total angular momentum on the molecular $z$ axis; $\varv$ is a generic vibrational quantum number/label, e.g. a combination  $(\varv_1,\varv_2,\varv_3)$ to describe vibrational states of a triatomic molecule; $g_{i}^{\rm ns}$ is the nuclear spin degeneracy; $T$ is the temperature, $c_2$ is the second radiation constant. In Eq.~(\ref{e:population}), $\tilde{E}_{J,k}^{\rm rot}$ is the rotational part of the ro-vibrational energy term value approximated as
\begin{equation}
\label{e:E=Evib+Erot}
\tilde{E}_{J,k,\varv} = \tilde{E}_{\varv}^{\rm vib} + \tilde{E}_{J,k}^{\rm rot},
\end{equation}
where $hc \cdot \tilde{E}_{\varv}^{\rm vib}$ is the vibrational ($J=0$) energy  (`band center') and $hc \cdot \tilde{E}_{J,k,\varv}$ is the total ro-vibrational energy. The non-LTE partition function $Q_{\rm nLTE}$ in Eq.~\eqref{e:population} is given by
\begin{equation}
\label{e:pf}
  Q_{\rm nLTE}(T) =\sum_{J,k,\varv}  g_{i}^{\rm ns}(2J+1) e^{-\frac{c_2 \tilde{E}_{J,k}^{\rm rot}}{T} } N_{\varv}^{\rm vib}.
\end{equation}

Our aim is to calculate the vibrational populations $N_{v} \equiv N_{v}^{\rm vib}$ of SiH\2\ as produced by instantaneous (vertical)  dissociation from three structures, \textit{d}GM, \textit{d}LM and \textit{d}TS. In case of \textit{d}LM and \textit{d}TS, the SiH$_2$ fragment can originate from any of the two different sides of their inverted structures, which should be taken into account. We therefore have to consider five different fragments, as shown in Table~\ref{t:fragments}.

\begin{table}
\caption{Five SiH\2\ fragments considered for the non-LTE analysis }
\label{t:fragments}
\begin{tabular}{ll}
\hline
\multicolumn{1}{c}{\sihh\ fragment name} & \multicolumn{1}{c}{Formation \& structure} \\
\hline
     GM\phantom{-R} &   From \textit{d}GM, two equivalent SiH\3\ \\
     TS-R & From \textit{d}TS, SiH\3\ pointing inside  \\
     TS-L & From \textit{d}TS, SiH\3\ pointing outside  \\
     LM-R & From \textit{d}LM, SiH\3\ pointing inside \\
     LM-L & From \textit{d}LM, SiH\3\ pointing outside  \\
\hline
\end{tabular}
\end{table}

Let us assume that the Si\2H\6\ is LTE and hence is in its ground vibrational state at the moment of dissociation, while the fragment SiH\2\ can end up in any vibrationally excited state $\varv \equiv (\varv_1,\varv_2,\varv_3)$ with some transition probability  giving rise to the vibrational population $N_{\varv}^{\rm vib}$. On top of that we also assume a full separation of the stretching Si--H and bending $\angle$HSiH modes inside Si\2H\6\ in its ground vibrational state. Possible consequences of deviation from these approximations are discussed below.

Under the assumptions made we define the vibrational population of the \sihh\ fragment as a Franck-Condon factor for a vertical transition from the ground vibrational state  of disilane Si$_{2}$H$_{6}$ to  SiH$_{2}$~+~SiH$_{4}$ with the gas phase (g.ph.) SiH\2\ transferred to some  vibrational state $\ket{\varv(\rm g.ph.)} = \ket{\varv_1,\varv_2,\varv_3}$. In the approximation of the full separation of the fragment SiH\2\ from the rest of \disil, the population of SiH\2\ can be represented as an overlap between the ground state wavefunction $\ket{\varv'' = 0({\rm fragment})}$ of a fragment \sihh\ and that of the corresponding vibrational state  $\ket{\varv' = \varv({\rm g. ph.})}$ of the gas phase \sihh\ as given by
\begin{equation}
\label{e:N=<0|v>^2}
N_{\varv} = |\langle 0 \, ({\rm fragment})  \ket{\varv\, ({\rm g.ph.})}|^2.
\end{equation}
Here the spacial wavefunction of the \sihh\ fragment  (i.e. a combination of two adjacent Si--H bonds in disilane with an angle between them forming the dissociating  \sihh) is projected on vibrational eigenfunctions of the gas phase \sihh\  to give the corresponding populations of \sihh.  The highest populated energy level will have the largest overlap between these wavefunctions.

The calculated temperature dependent populations $N_{J,k,\varv}(T)$ in Eq.~\eqref{e:population} can be then combined with a molecular line list for SiH\2\ to simulate absorption or emission spectra of this molecule under the non-LTE conditions as defined by $N_{J,k,\varv}(T) $ in Eq.~\eqref{e:population}. Here we use the ExoMol line list CATS  by \citet{jt779}  as provided by ExoMol (\url{www.exomol.com}).
Technically this is done by incorporating the non-LTE vibrational densities $N_{\varv}$ into the ExoMol States file as described in Section \ref{sec:comp} (the ExoMol file formats are discussed extensively elsewhere \citep{jt810}).  A non-LTE spectrum of \sihh\ for given $T$ and $P$ is then calculated using CATS' Einstein-$A$ coefficients with the \xc\ program,\citep{ExoCross} where a new non-LTE option has been implemented as part of this work. The rotational populations are assumed to be in LTE according with Eq.~\eqref{e:population}.

\section{Computing vibrational populations of SiH\2}
\label{sec:comp}

Two approaches were used for the calculation of the population densities of the fragment \sihh.  One approach - named the decoupled 1D approach - is where the 3D wavefunctions of the fragment as well as of free \sihh\ are represented by products of 1D parts with the harmonic oscillators as wavefunctions. This simplified  model is mainly used to illustrate the idea of our non-LTE treatment.
The second, more accurate approach - named the 3D approach - is based on the full 3D vibrational wavefunctions computed using the variational program  \trove.\citep{TROVE} Both approaches are presented in the following in order to assess and compare the accuracy achieved.

\subsection{The 1D approach}
\label{sec:1d-approach}

A vibrational state $\ket{\varv}=\ket{\varv_1,\varv_2,\varv_3}$ of \sihh\ is characterized by the three (normal mode) quantum numbers $\varv_1$, $\varv_2$ and $\varv_3$ corresponding to the two stretching modes ($\varv_1$ and $\varv_3$) and one bending mode ($\varv_2$) of the \sihh\ molecule.  The 1D approach considers the stretching Si--H$_1$, Si--H$_2$ and bending $\angle$HSiH modes,  both of the molecular fragment and gas phase \sihh\ molecules, as fully independent and described by one dimensional (1D) wavefunctions under the harmonic approximation, as given by:
\begin{equation}
    \label{e:harmonic}
 \Psi_v(x) = C_\varv\, H_\varv(x)\, e^{-\frac{x^2}{2}}.
\end{equation}
Here $x$ is a dimensionless coordinate describing either the stretching $r = r_{\rm Si-H}$ or bending $\alpha=\alpha_{\angle{\rm H-Si-H}}$ coordinate as follows:
\begin{eqnarray}
x_{\rm s} &=& \frac{r-r_{\rm e}}{\sqrt{a_{\rm s}}}, \\
x_{\rm b} &=& \frac{\alpha-\alpha_{\rm e}}{\sqrt{a_{\rm b}}},
\end{eqnarray}
with
\begin{eqnarray}
a_{\rm s} &=& \frac{h}{4\pi^2 c \omega_{\rm s}}  \left[ \frac{1}{M_{\rm Si}} + \frac{1}{M_{\rm H}} \right], \\
a_{\rm b} &=& \frac{h}{4\pi^2 c  \omega_{\rm b}} \left[ \frac{2(1- \cos\alpha_{\rm e})}{M_{\rm Si} {r_{\rm e}^{2}}}+\frac{2}{M_{\rm H} r_{\rm e}^{2}} \right]
\end{eqnarray}
and  $\omega_{\rm s} = \omega_{\rm Si-H}$,  $\omega_{\rm b} = \omega_{\angle{\rm HSiH}}$. In Eq.~\eqref{e:harmonic} $H_\varv(x)$ is a Hermite polynomial and $C_\varv$ is the corresponding normalization constant.

The constants $a_{\rm s}$ and $a_{\rm b}$ correspond to inverse masses of the vibrational part of a free three-atomic molecule expressed in terms of the internal coordinates $r_1$, $r_2$, $\alpha$ (see, e.g. \citet{jt96,20YuMeFr}).

A 1D population for the corresponding mode of the gas phase SiH\2\ molecule is  given by Eq.~\eqref{e:N=<0|v>^2} with
$\ket{\varv} = \Psi_\varv(x)$.
The different disilane fragments have different structural parameters $r_{\rm e}$, $\alpha_{\rm e}$, $\omega_{\rm s}$ and $\omega_{\rm b}$, see Tables~\ref{t:structures} and \ref{t:freq-2}, and thus lead to different ground state vibrational 1D wavefunctions $\ket{0 ({\rm fragment})}$  (stretching or bending) and hence result in different vibrational populations $N_\varv$ of the gas phase SiH\2\ according with Eq.~\eqref{e:N=<0|v>^2}.

For each of the three modes (two stretching and one bending),  1D wavefunctions of the gas phase \sihh\ for 30 vibrational states from $\varv_{0}$ up to $\varv_{29}$ were  calculated.  These 1D wavefunctions were then numerically integrated with the corresponding ground state  1D wavefunctions $\ket{\varv=0}$ of the fragment in question.

The total vibrational population $N_{\varv_1,\varv_2,\varv_3}$ in this approximation is then given by a product
\begin{equation} \label{eq:density-3D}
N_{\varv_1,\varv_2,\varv_3} = N_{\varv_1}^{\rm str} N_{\varv_2}^{\rm bnd} N_{\varv_3}^{\rm str}
\end{equation}
where $N_{\varv_1}^{\rm str}$ and $N_{\varv_3}^{\rm str}$ are obtained using the stretching harmonic oscillators wavefunctions $\ket{\varv_1}$ and $\ket{\varv_3}$, while $N_{\varv_2}^{\rm bnd}$ is obtained using the corresponding bending harmonic oscillator wavefunction $\ket{\varv_2}$.  The independent treatment of the two stretching populations is partly justified by the local mode character of the vibrational degrees of freedom of SiH\2\ due to the 90$^{\circ}$ bond angle (see, e.g. \citet{00Jensen.cluster} and \citet{jt779}). The asymmetric vibrational modes of SiH\2\ ($B_2$ in \Cv{2}) are non populated  in this 1D approximation. This is because for  the parallel nature of the Franck-Condon transitions from the ground vibrational state of disilane, which  is fully symmetric ($A_1$) and the excitation, only symmetric states of SiH\2\ give rise to non-zero integrals in Eq.~\eqref{e:N=<0|v>^2}. For example, the vibrational population of the (2,1,0) state  (2 stretching and 1 bending quanta of $A_1$) of the GM fragment is obtained as a product of $N_{\varv_1}^{\rm str} = $ 0.274, $N_{\varv_2}^{\rm bnd} = $  0.333 and $N_{\varv_3}^{\rm str}= $ 0.967, respectively, resulting in $N_{(2,1,0)}$ = 0.088, while the population of the $B_2$-type  (0,0,1) vibratitonal state is assumed to be zero. The populations $N_{\varv_1,\varv_2,\varv_3}$  are  pre-calculated for each vibrational state $(\varv_1,\varv_2,\varv_3)$ of SiH\2\ and added to the CATS State file to be used in non-LTE simulations (see below, Section~\ref{sec:non-LTE-simulations}).
All the vibrational populations computed and used as part of this work are provided in the supplementary material.

\subsubsection{1D populations and spectra}

\begin{figure*}[ht]
\centering
\includegraphics[width=0.6\columnwidth]{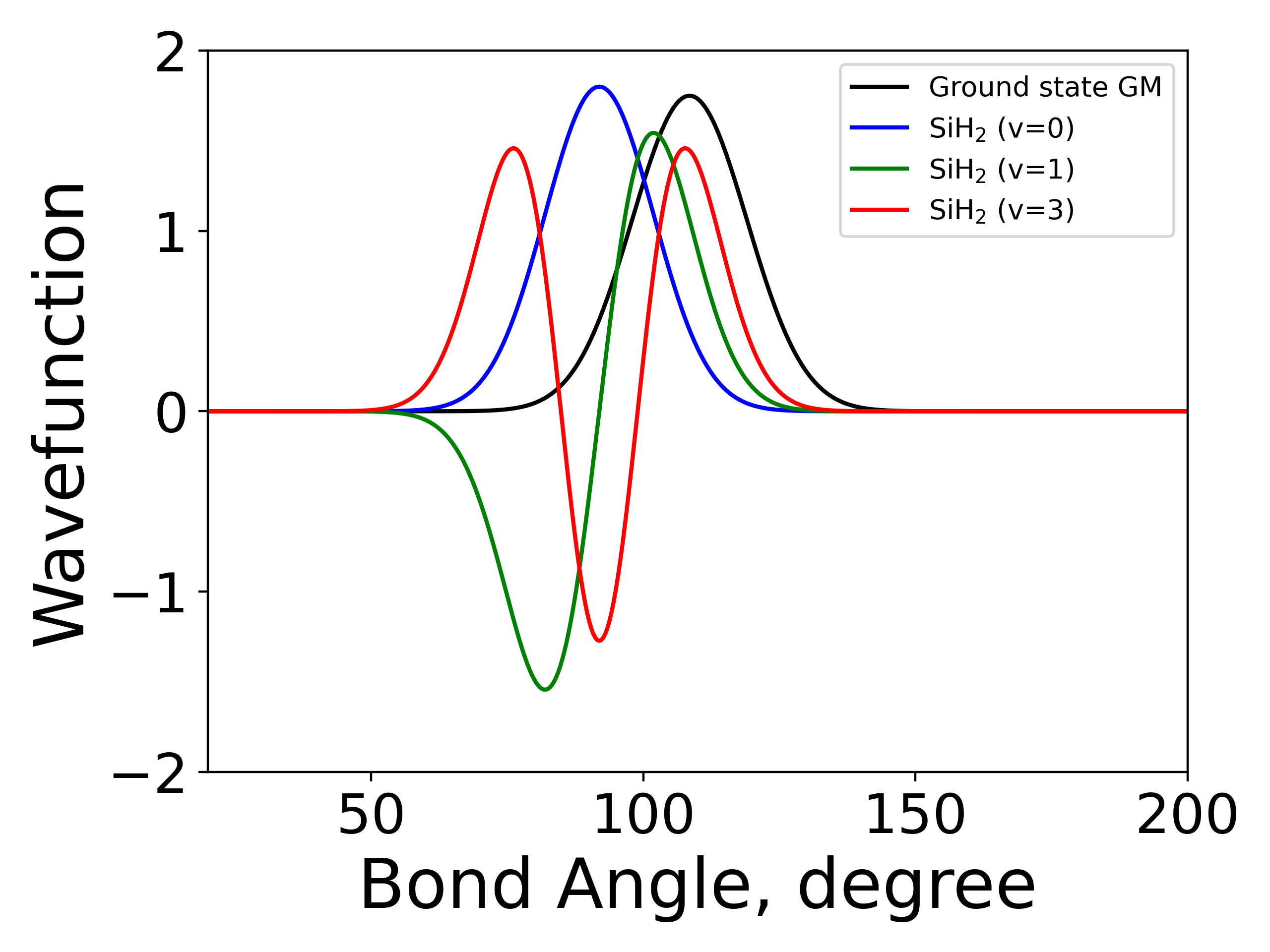}
\includegraphics[width=0.6\columnwidth]{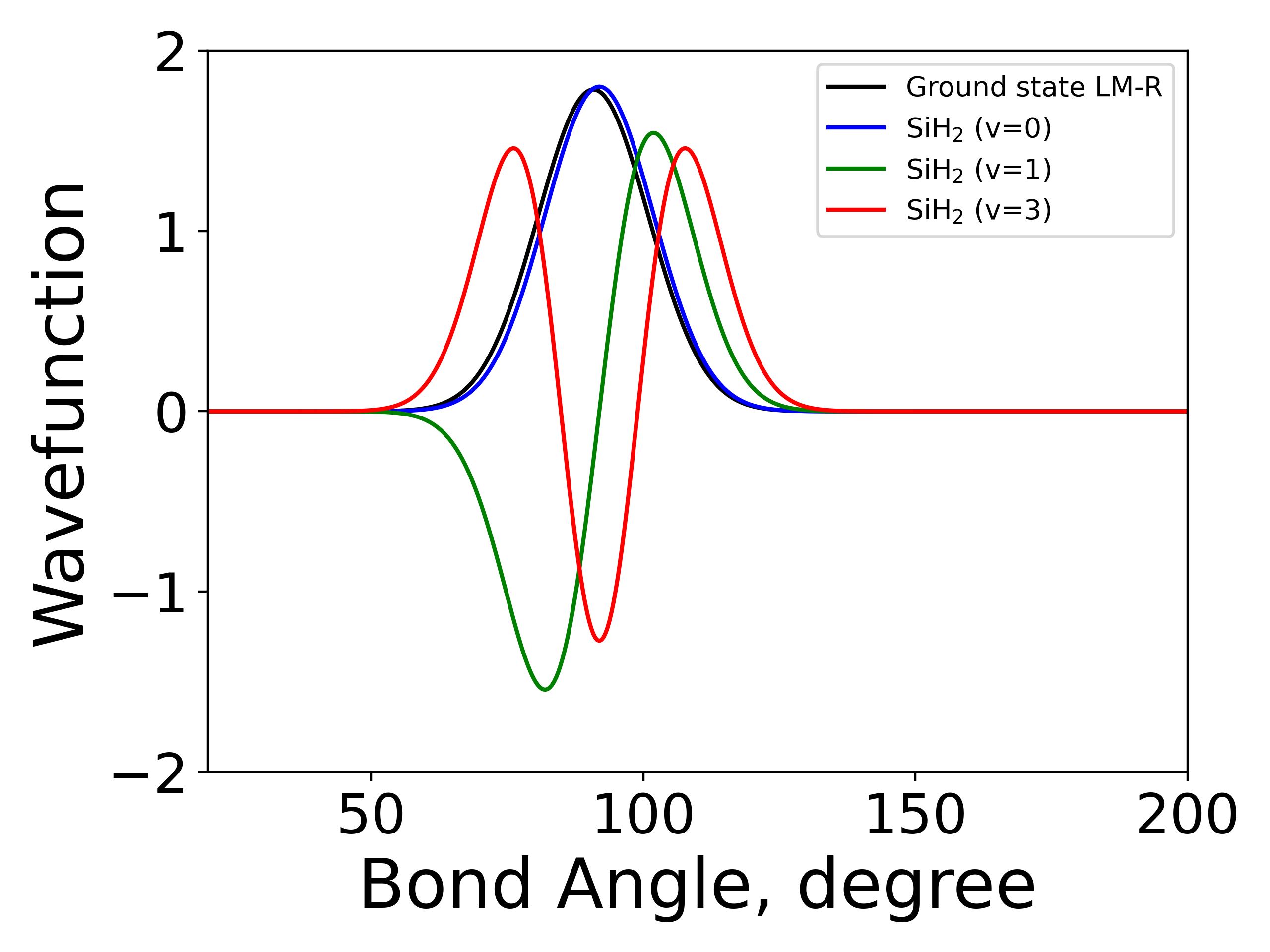}
\includegraphics[width=0.6\columnwidth]{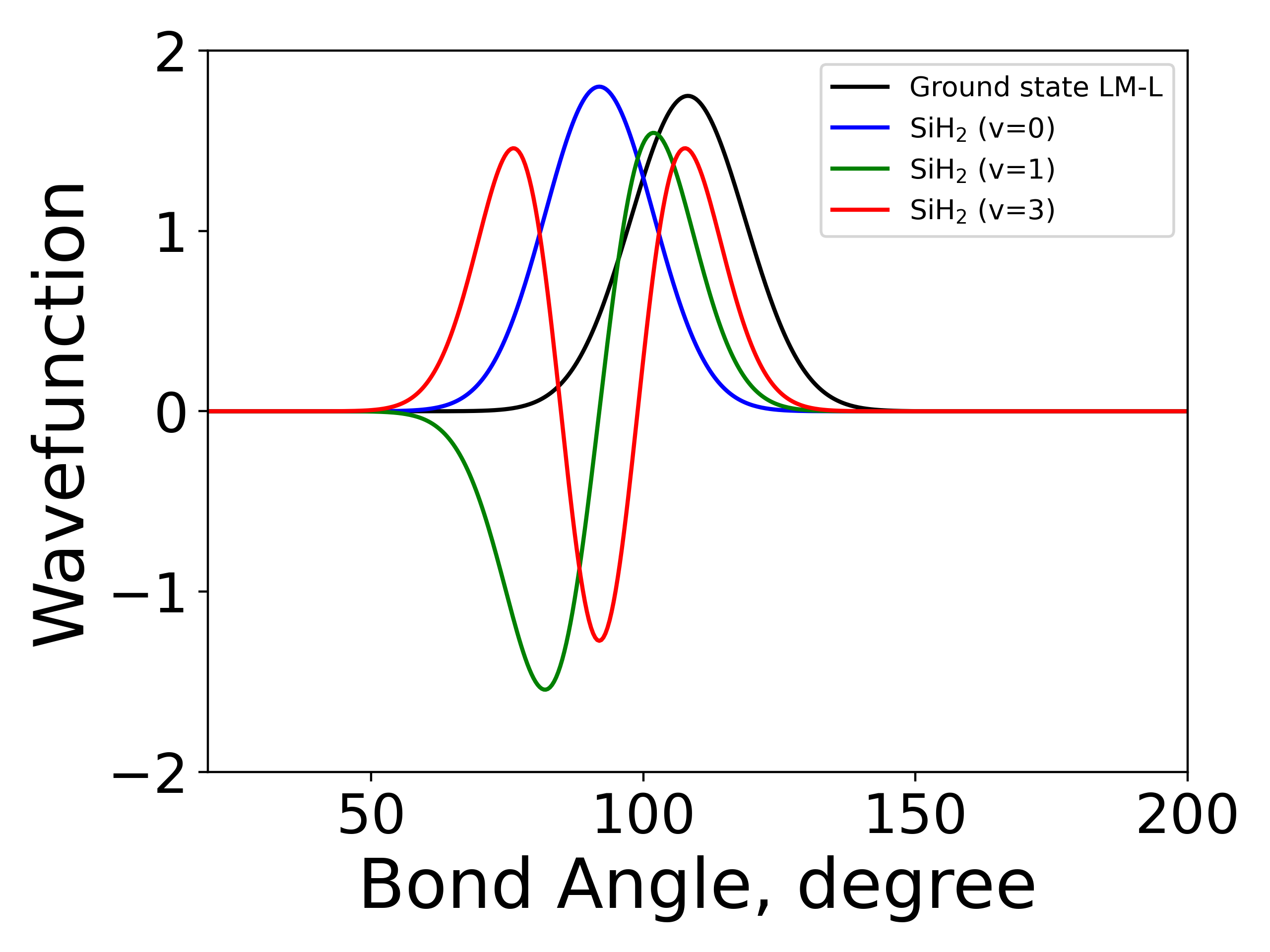}
\includegraphics[width=0.6\columnwidth]{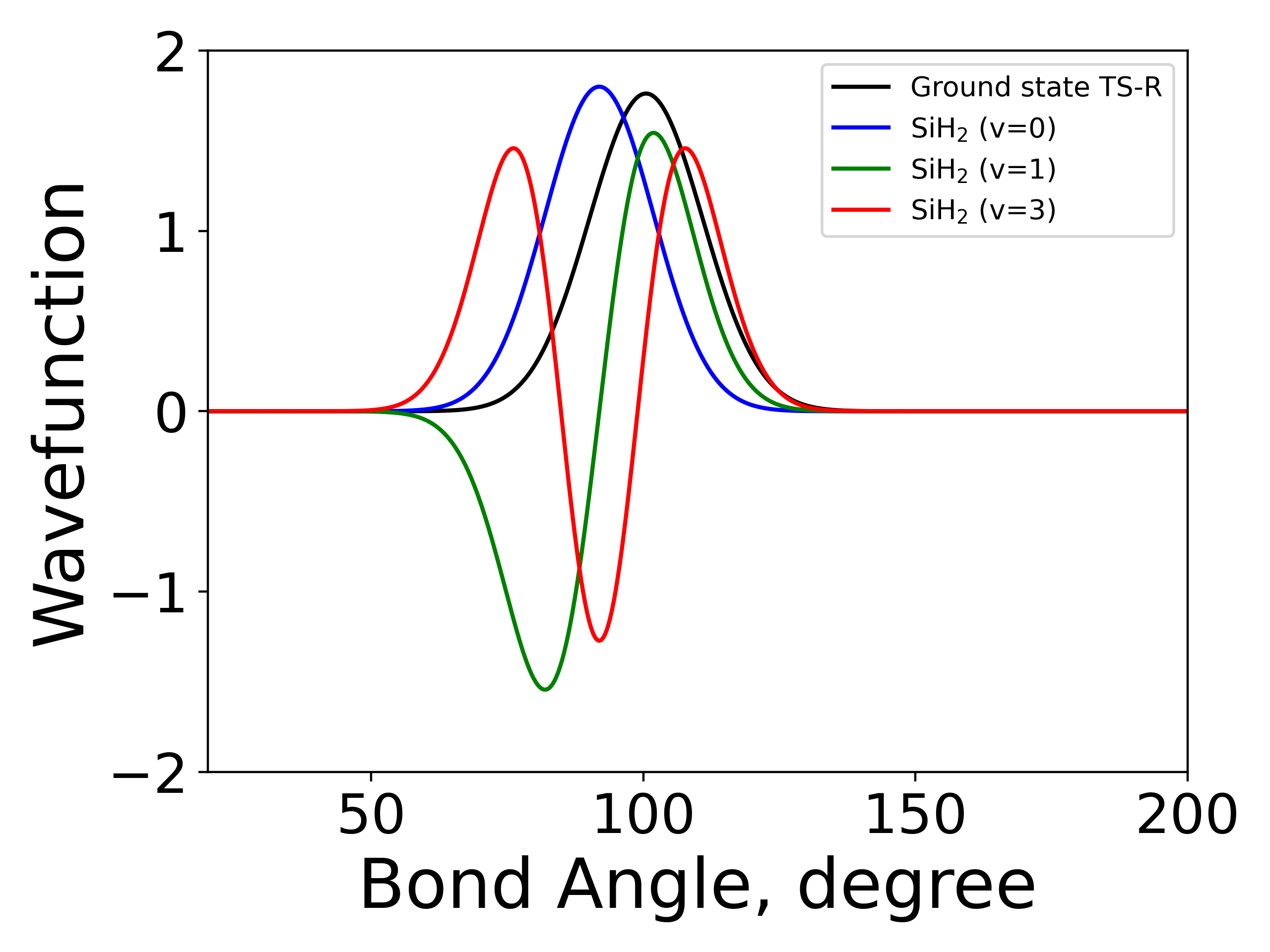}
\includegraphics[width=0.6\columnwidth]{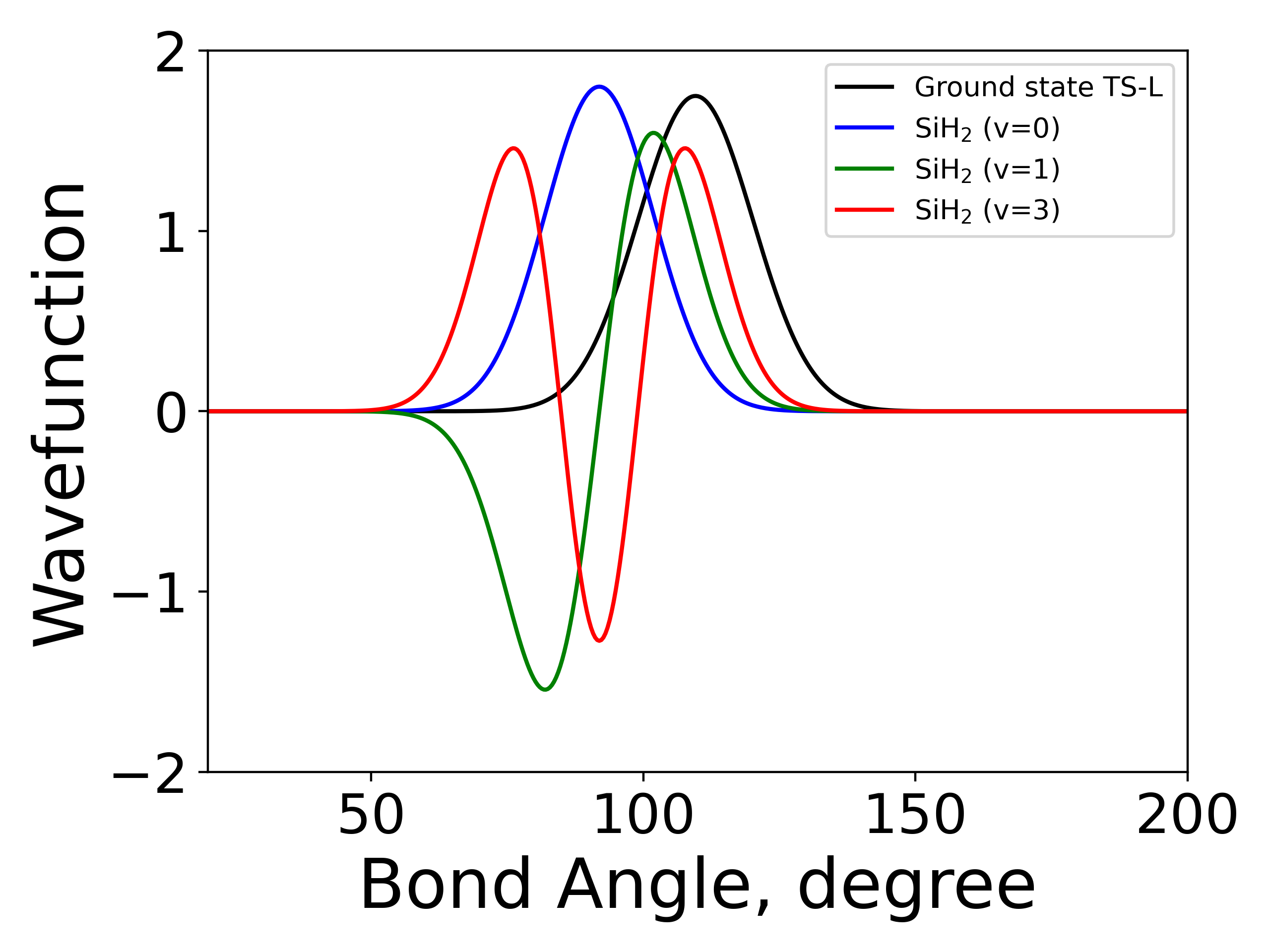}
\caption{
The 1D bending mode harmonic wavefunctions for the five \sihh\ fragments described in Table \ref{t:structures} as a function of bond angle in degrees. Ground state fragment (black) compared with the LTE $\ket{\varv=0}$ (blue), $\ket{\varv=1}$ (green), $\ket{\varv=2}$ (red)  }
\label{fig:1doverlapsbend}
\end{figure*}

\begin{figure*}[ht]
\centering
\includegraphics[width=0.6\columnwidth]{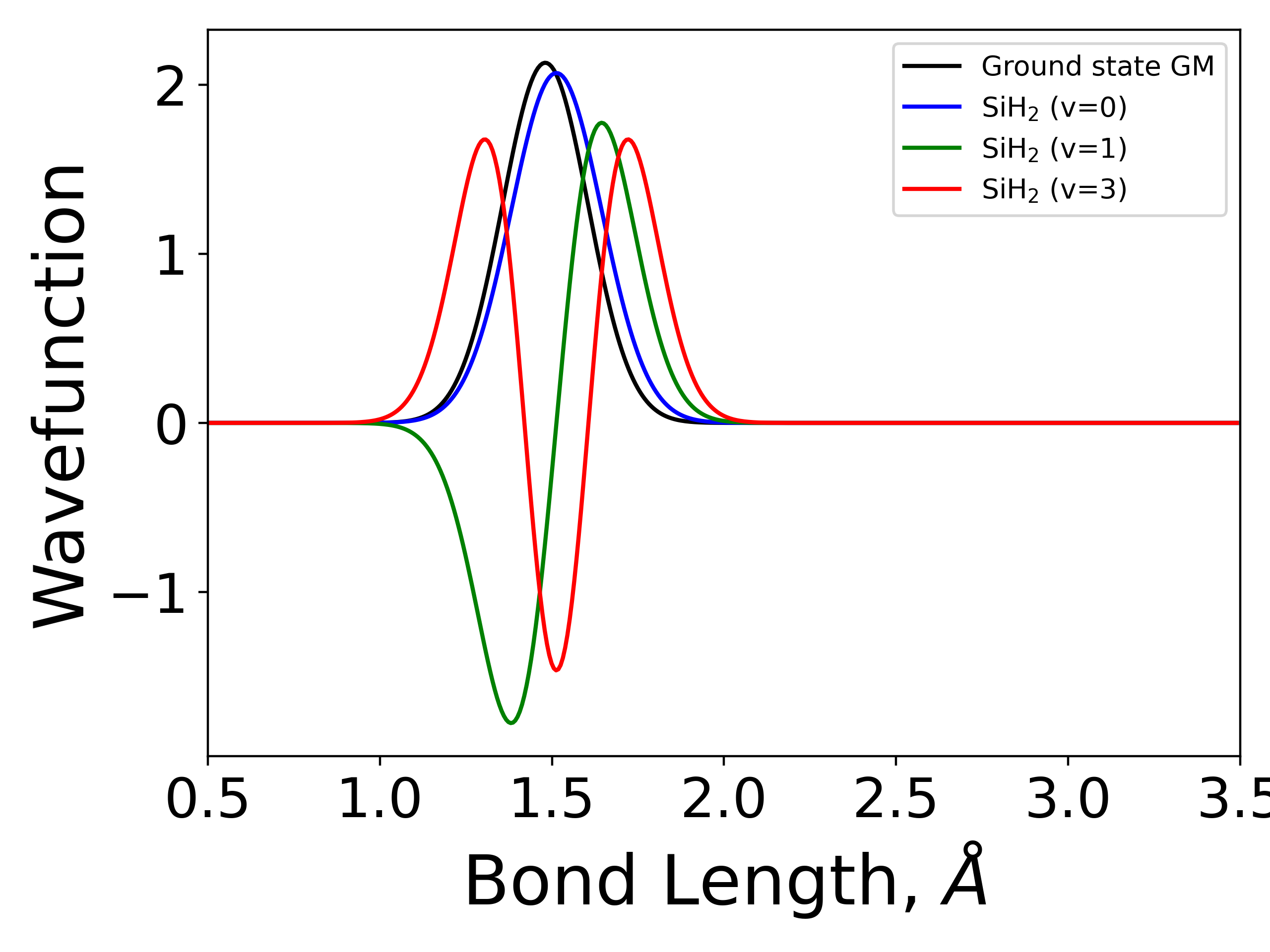}
\includegraphics[width=0.6\columnwidth]{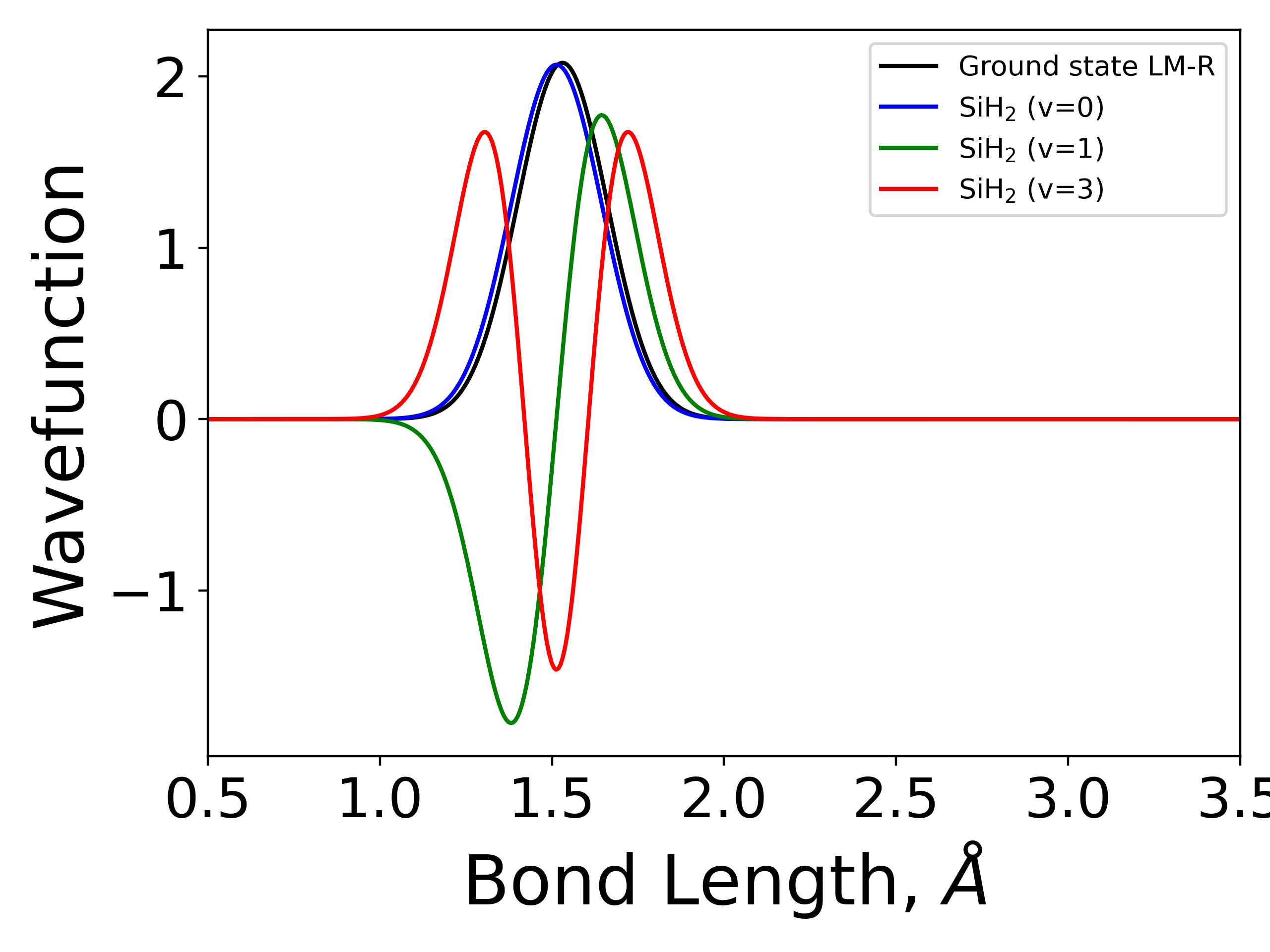}
\includegraphics[width=0.6\columnwidth]{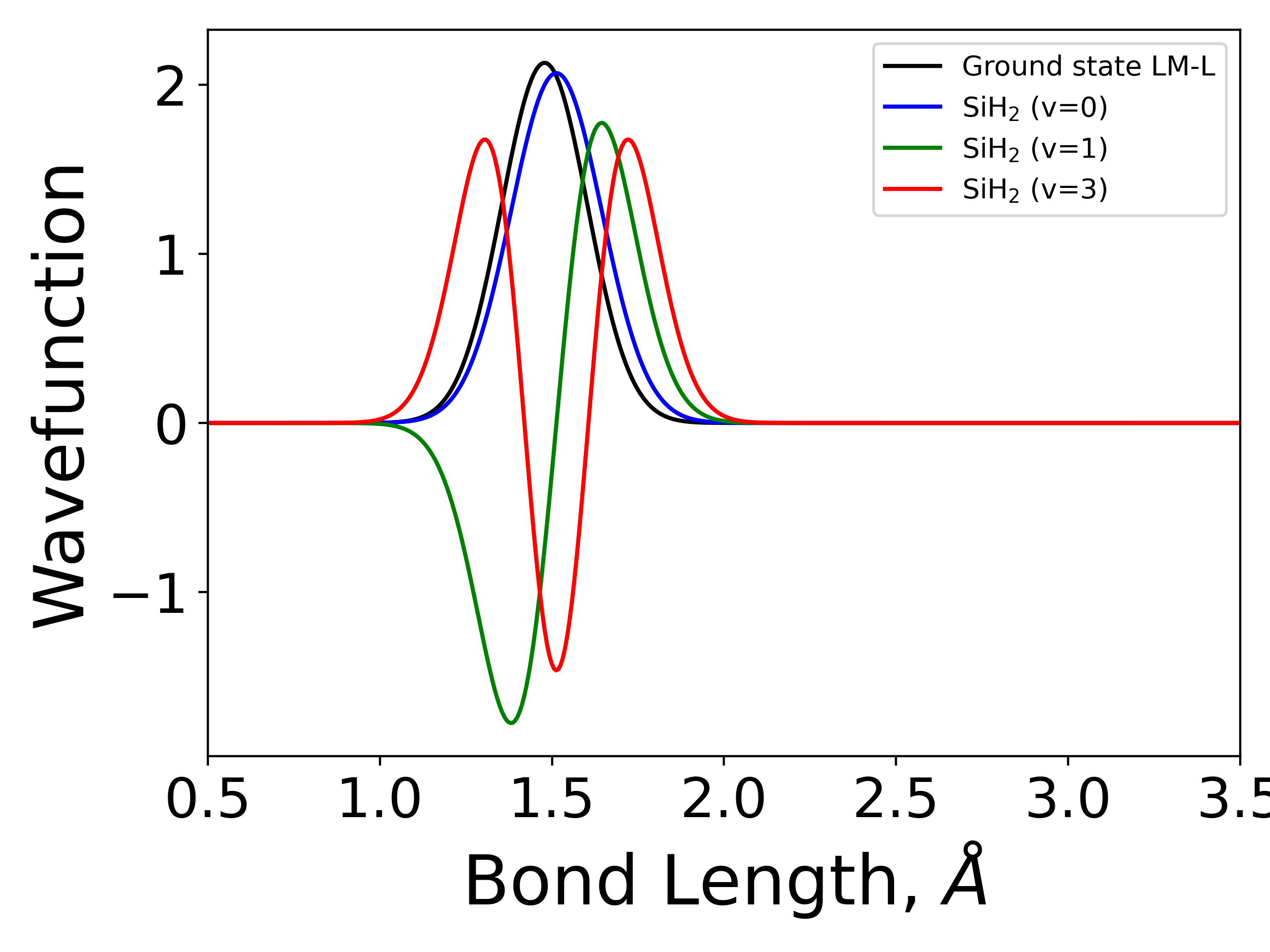}
\includegraphics[width=0.6\columnwidth]{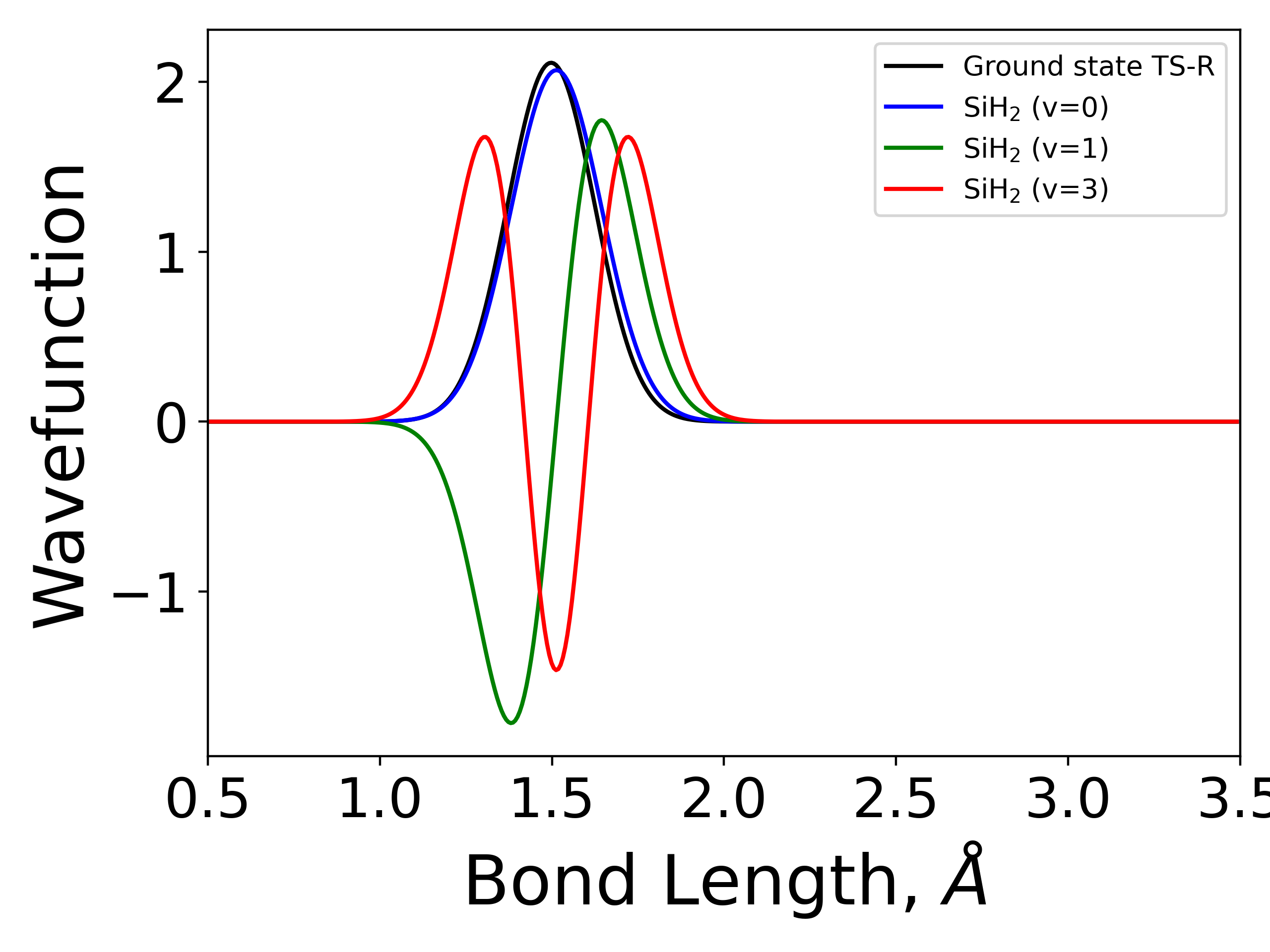}
\includegraphics[width=0.6\columnwidth]{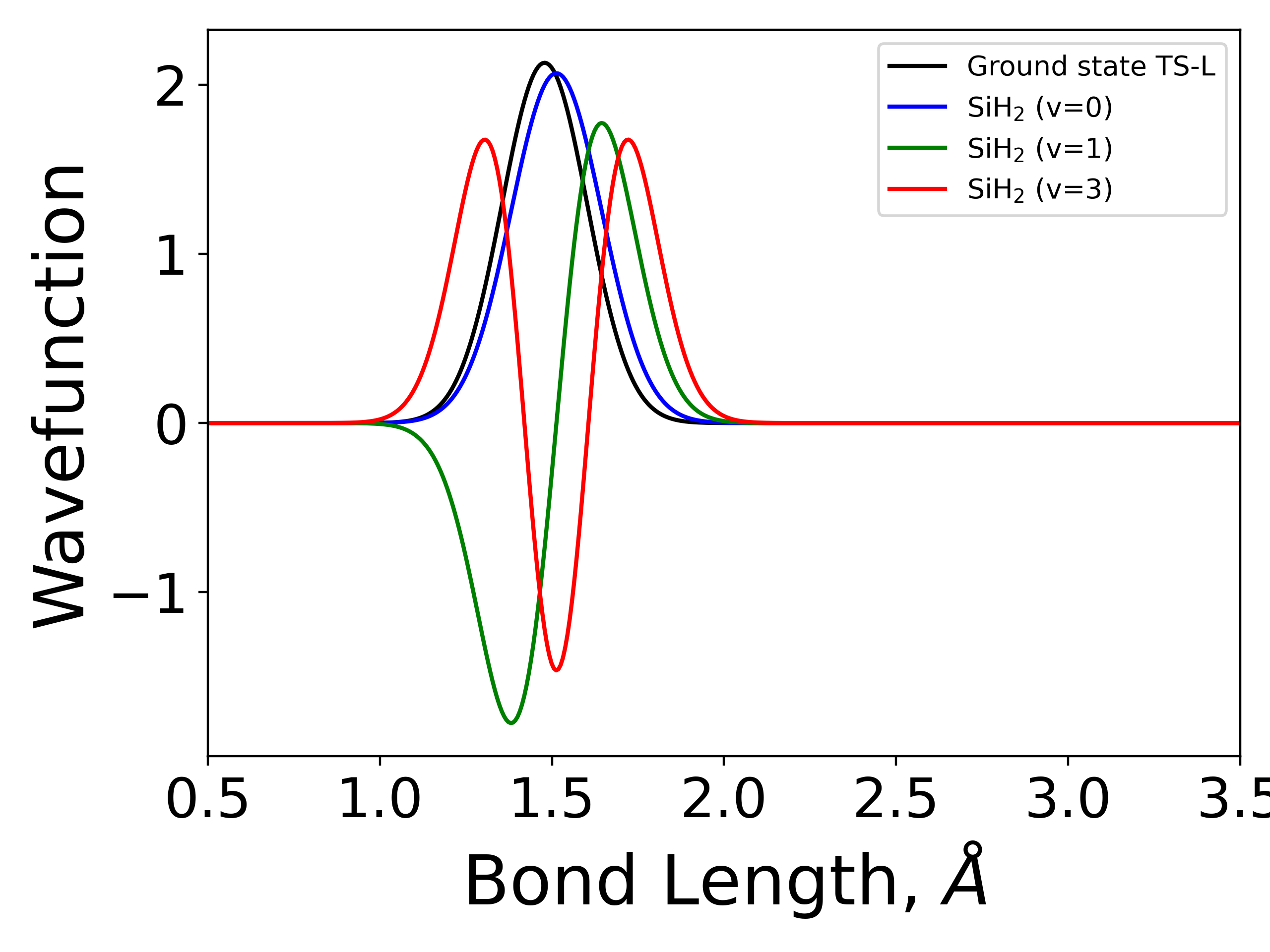}
\caption{
The 1D stretching mode harmonic wavefunctions for the five \sihh\ fragments described in Table \ref{t:structures} as a function of bond length in Angstrom. Ground state fragment (black) compared with the LTE $\ket{\varv=0}$ (blue), $\ket{\varv=1}$ (green), $\ket{\varv=2}$ (red)
}
\label{fig:1doverlapsstretch}
\end{figure*}

\begin{figure}[ht]
\centering
\includegraphics[width=0.6\columnwidth]{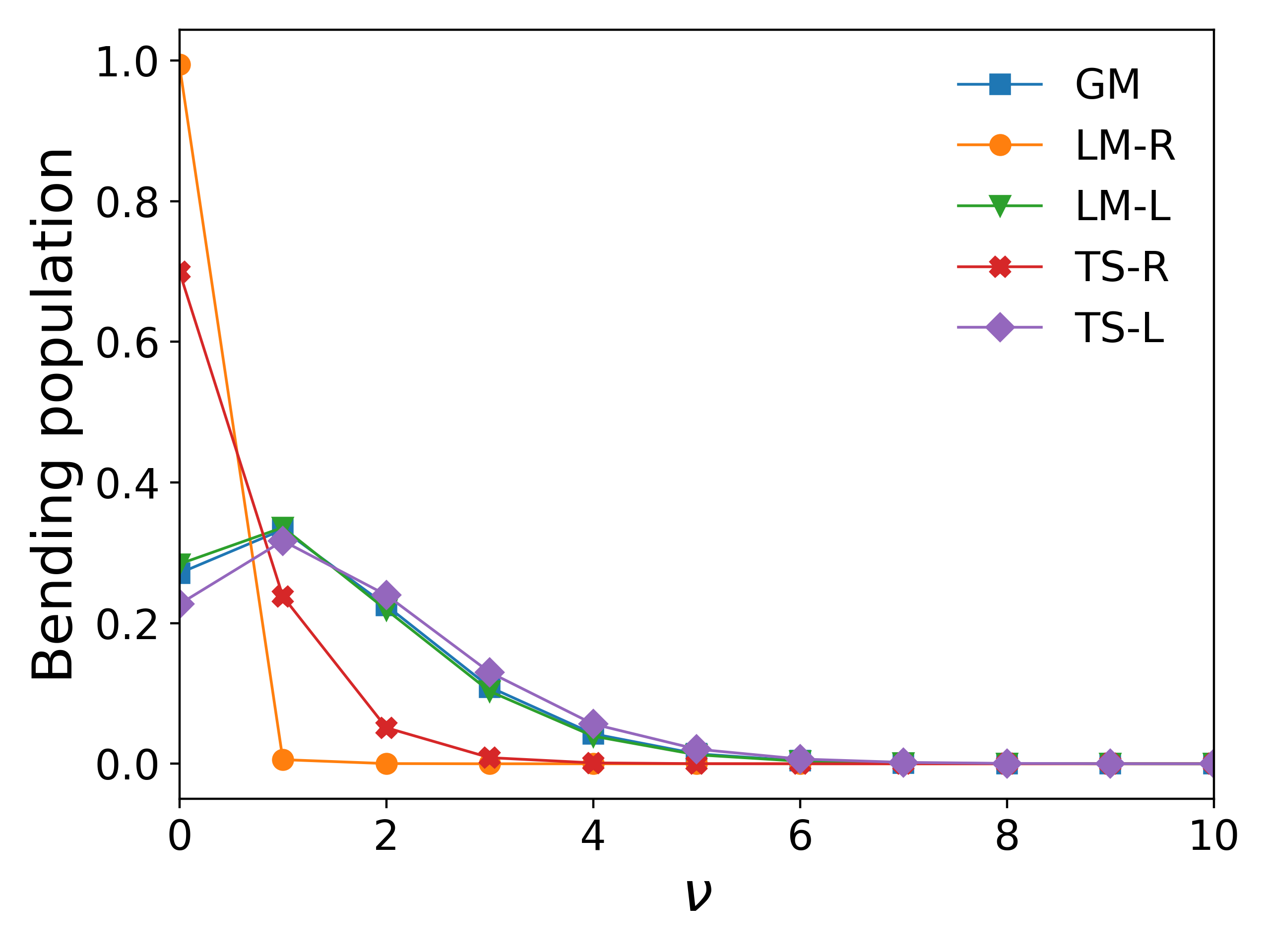}
\includegraphics[width=0.6\columnwidth]{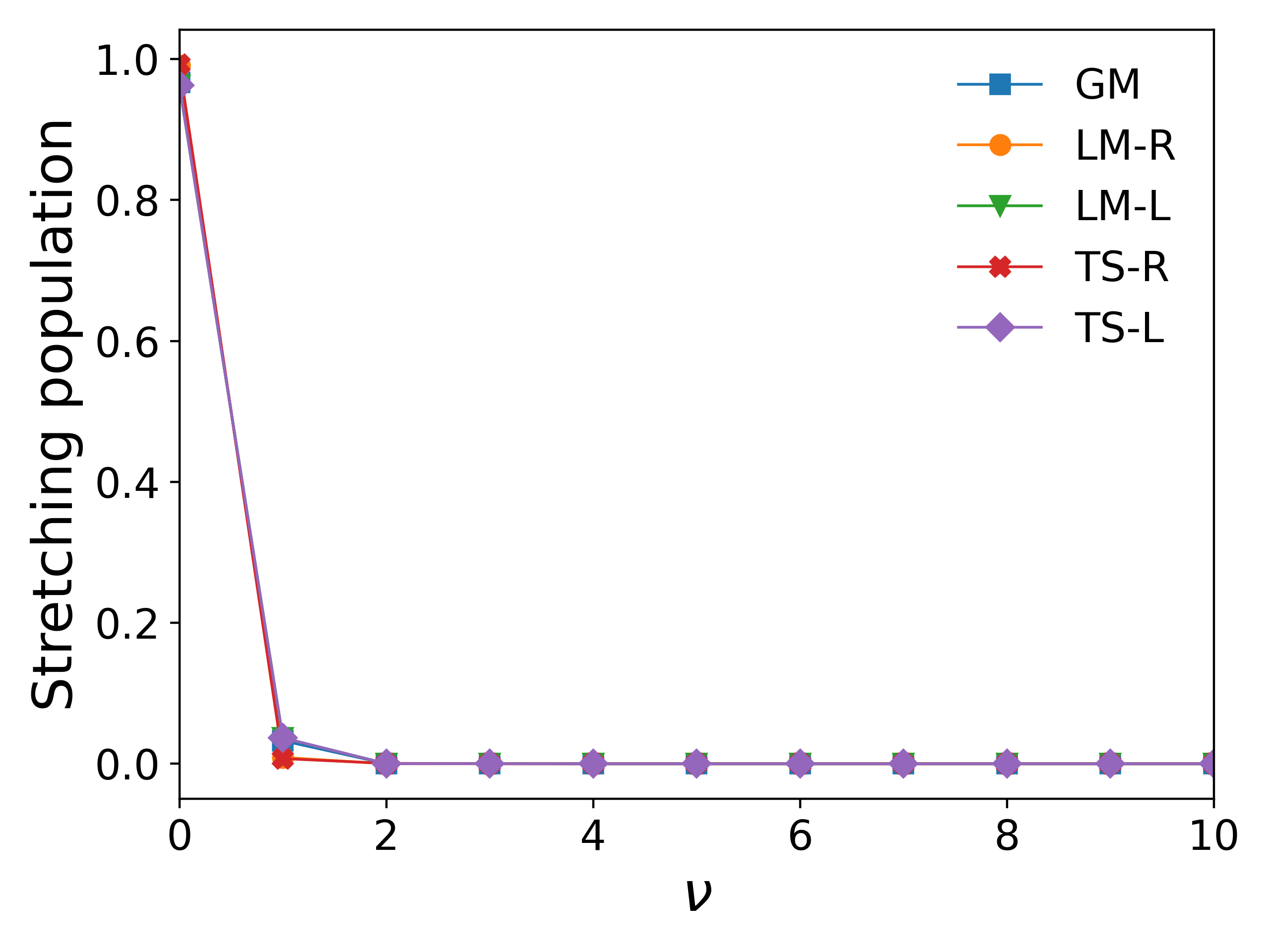}
\caption{Stretching and bending mode 1D populations $N_{v}$ for the five \sihh\ fragments from $\varv' = 0$ -- $\varv'=10$ }
\label{fig:2d-populations-python}
\end{figure}

Examples of  overlapping bending mode Harmonic wavefunctions used in calculations of populations for $\varv=0, 1$ and 2 of the five fragments are shown in Figures \ref{fig:1doverlapsbend} and \ref{fig:1doverlapsstretch} for the bending and stretching modes respectively.  In all cases the black curve represents the ground state wavefunction for the non-LTE fragment, and the blue, green and red line show the $\varv={0}$, $\varv={1}$ and $\varv={2}$ LTE wavefunctions of gas phase \sihh.\citep{jt779}  The corresponding 1D populations  as an integral of the overlaps between the LTE and non-LTE wavefunctions  are  plotted in Figure \ref{fig:2d-populations-python} for the bending and stretching modes of the five fragments. The stretching populations exhibit a typical Boltzmann-like distribution with the ground vibrational state $\varv=0$ as the mostly populated in all five cases. This is expected because their equilibrium bond lengths are rather similar to that of the gas phase SiH\2. In case of the bending populations, only   LM-R and TS-R have $\varv=0$ to be with the highest populations, while for LM-L, TS-L and GM the $N_{\varv}$  distributions exhibit strong non-LTE character  with $\varv=2$ to be almost as populated as the ground vibrational state $\varv=0$.

The shapes and positions of the curves in Figures~\ref{fig:1doverlapsbend} and \ref{fig:1doverlapsstretch} match with the parameters  from Table~\ref{t:structures}.  The larger $\omega$ used for the bending modes manifests itself as wider curves, while the black curves are all centred around the equilibrium bond angles and length listed  in Table~\ref{t:structures}.

The similarity between the populations of the LM-L, TS-L and GM fragments is expected owing to their similar structural parameters.  It is interesting to see the most populated vibrational levels of LM-R and TS-R are always lower than the vibrational levels of LM-L, TS-L and GM.


\begin{figure*}[ht]
\centering
\includegraphics[width=0.6\columnwidth]{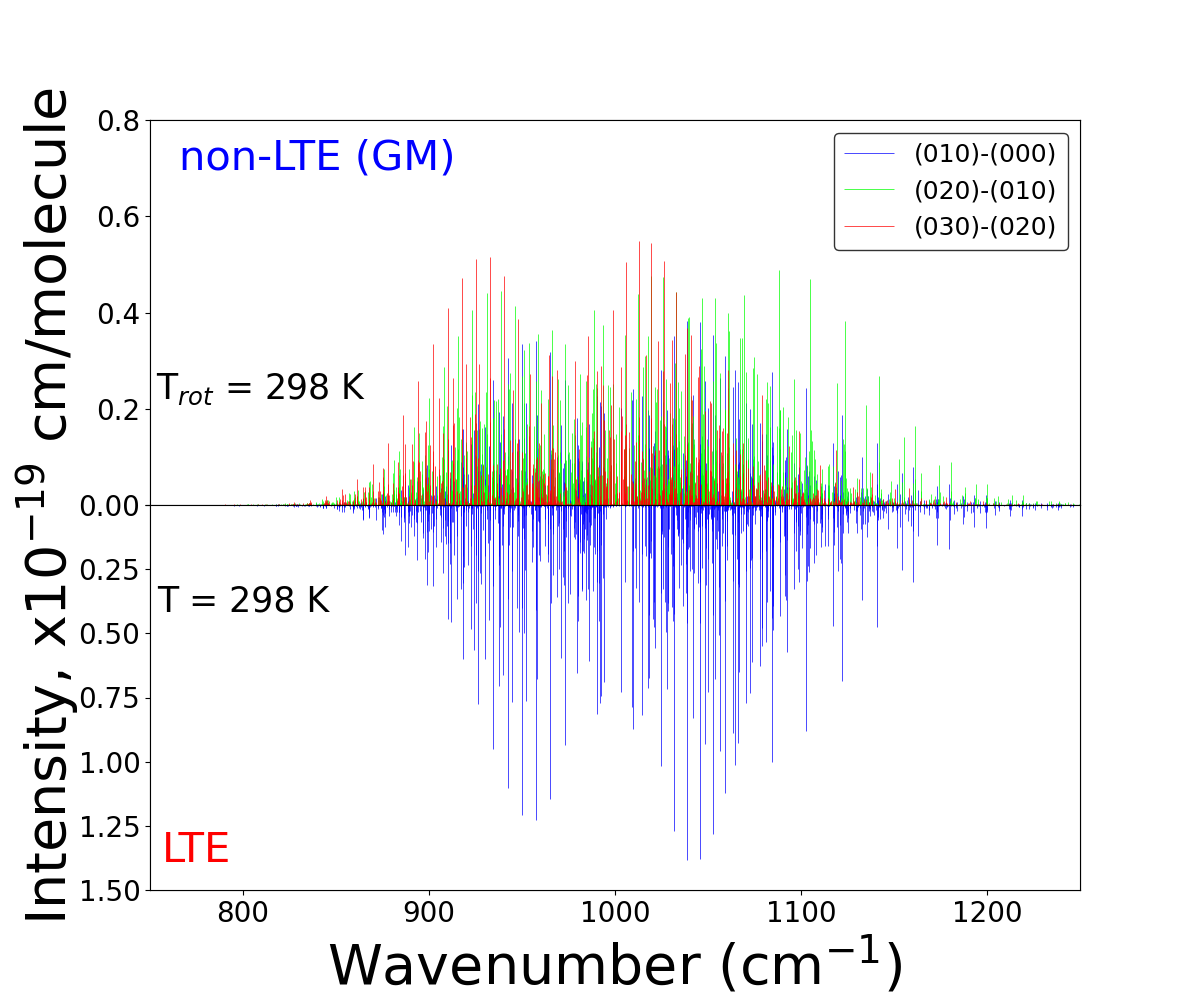}
\includegraphics[width=0.6\columnwidth]{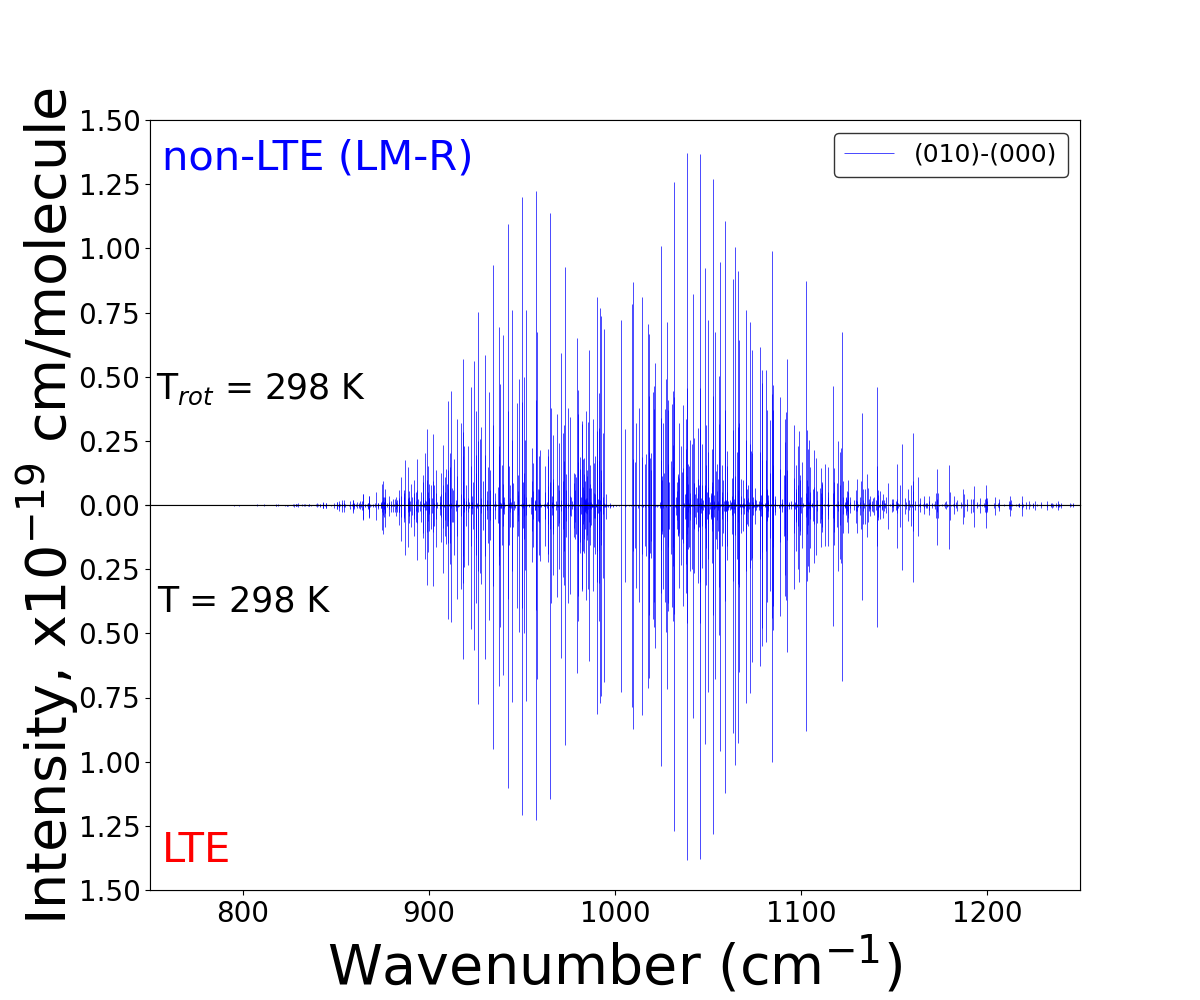}
\includegraphics[width=0.6\columnwidth]{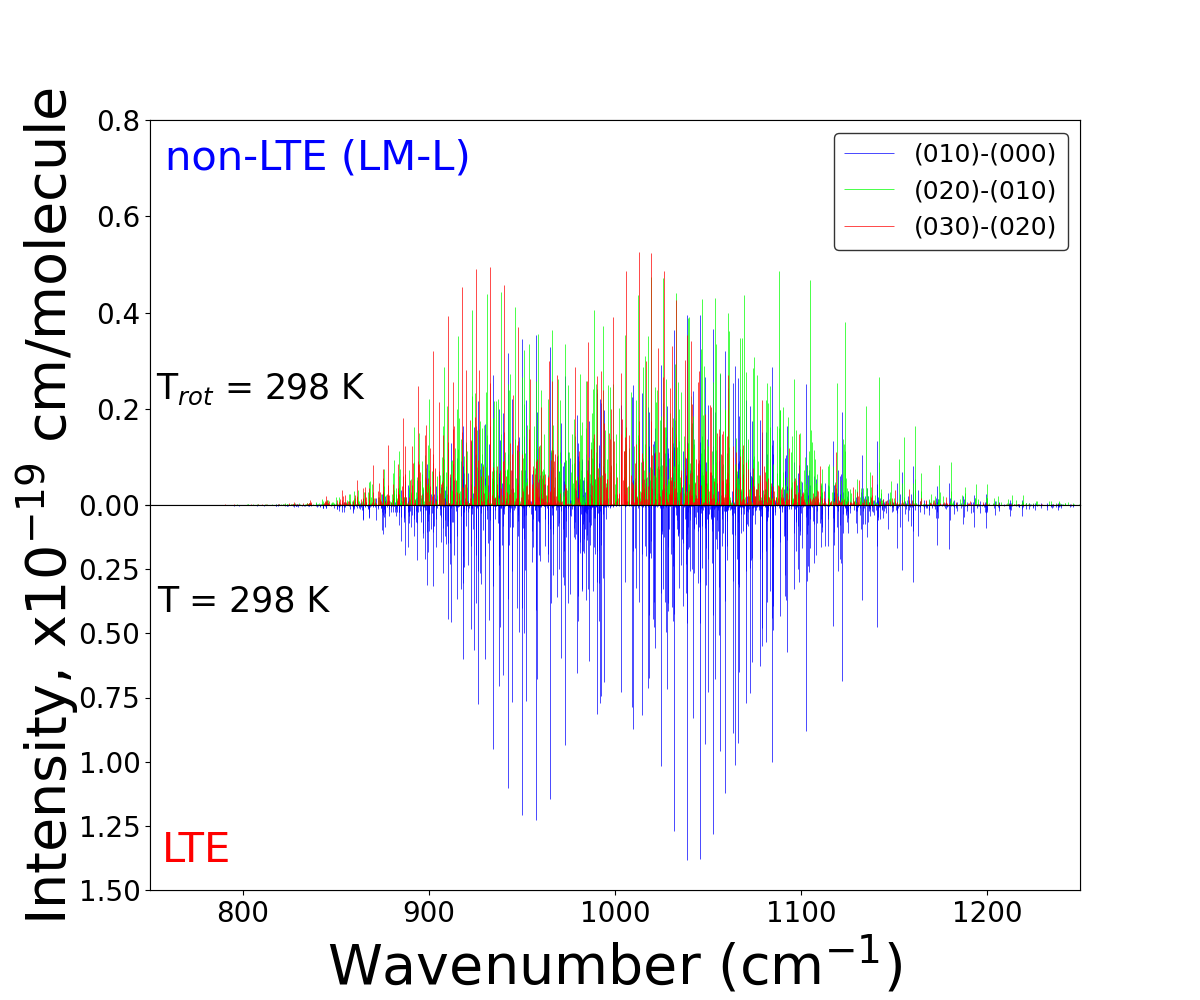}
\includegraphics[width=0.6\columnwidth]{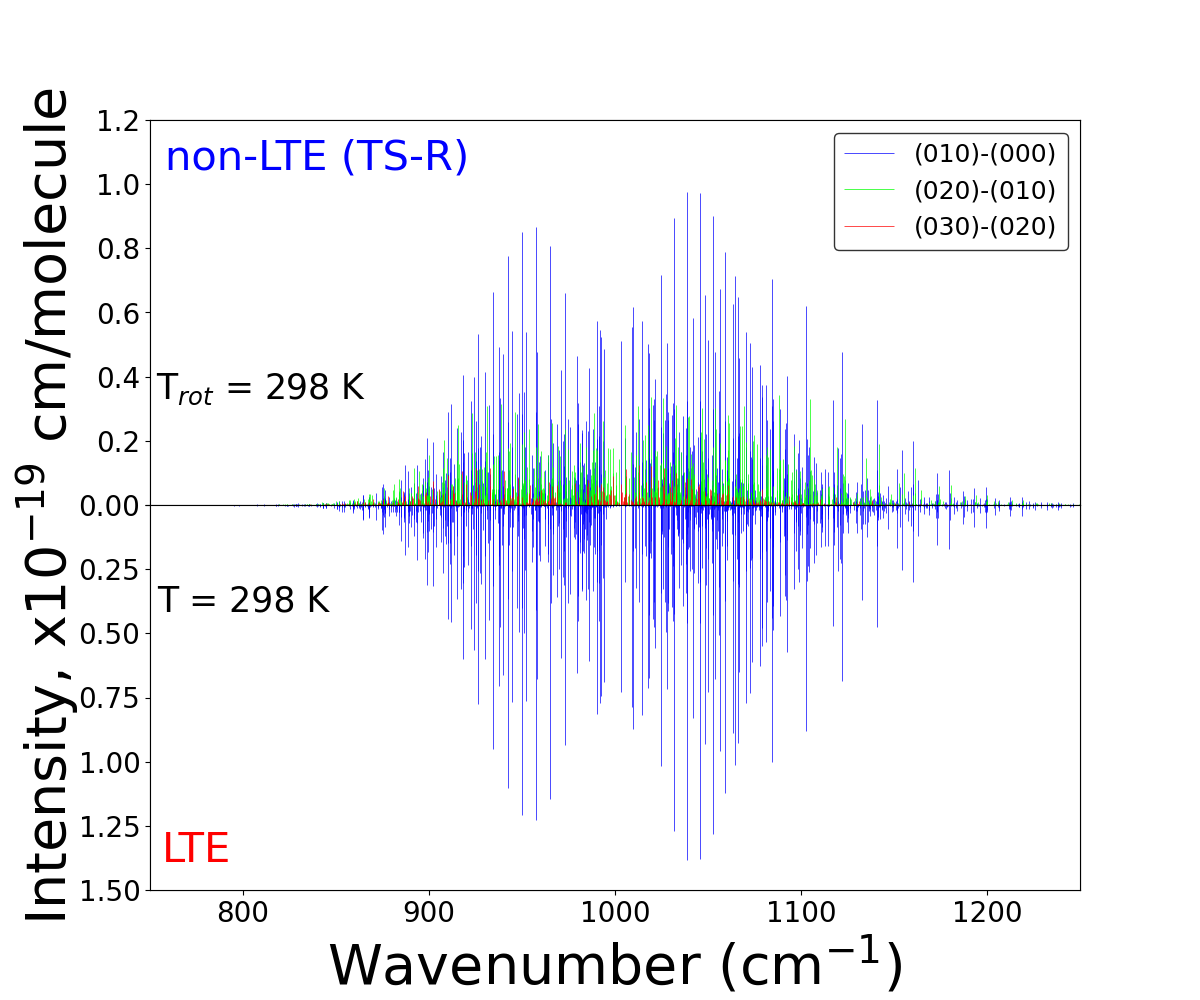}
\includegraphics[width=0.6\columnwidth]{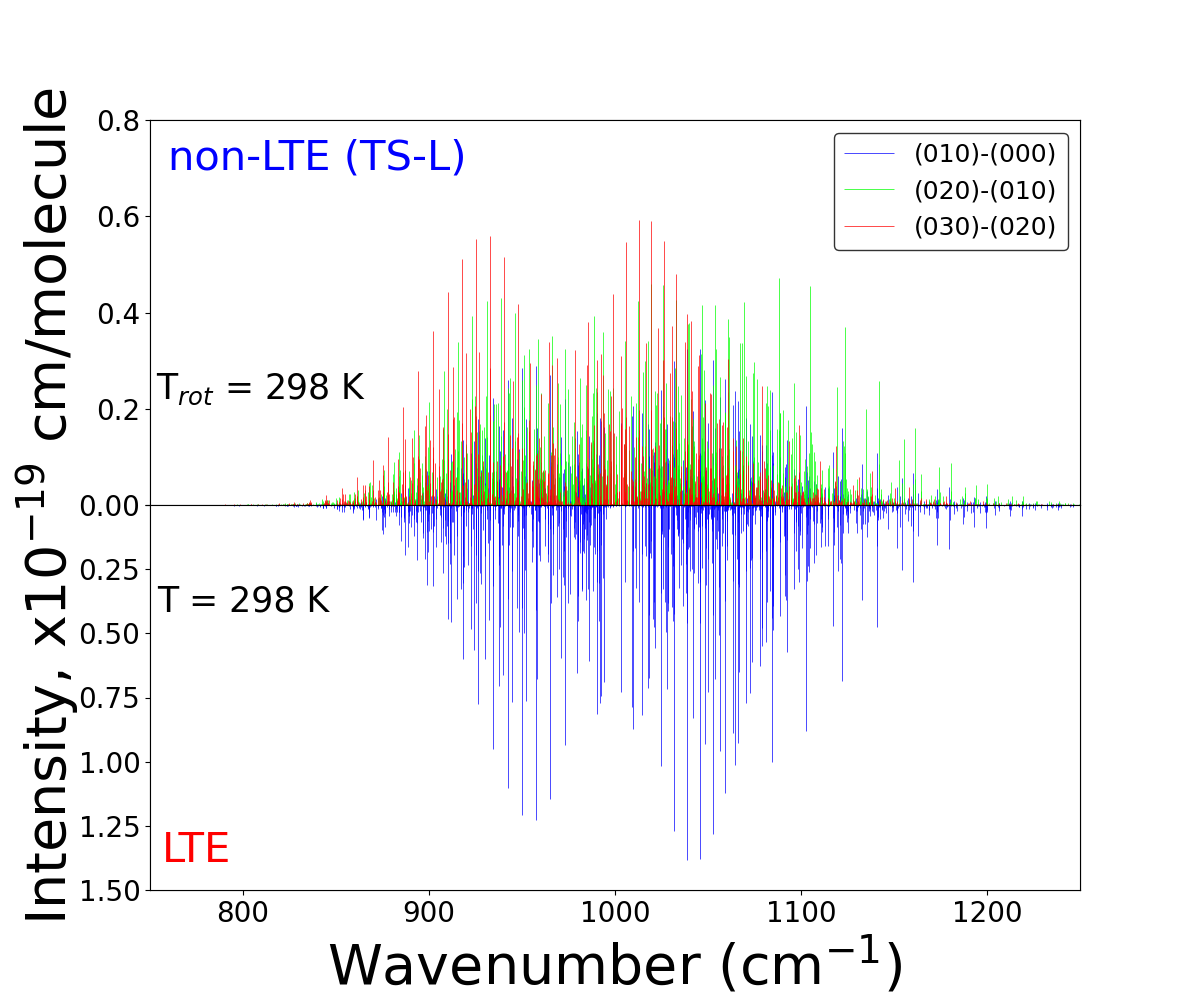}
\caption{Non-LTE spectra of SiH\2\ at $T_{\rm rot}$ = 296~K in the 1000~\icm\ (10~$\mu$m) region corresponding to five  vibrational populations GM, TS-L, TS-R, LM-L and LM-R (upper displays of each figure) and compared to the same spectra simulated using the LTE population at $T=296$~K. The non-LTE populations were obtained using the 1D approach (see text). Only the strongest bands are shown. The CATS line list was used }
\label{fig:filter-1d-1000}
\end{figure*}

\begin{figure*}[ht]
\centering
\includegraphics[width=0.6\columnwidth]{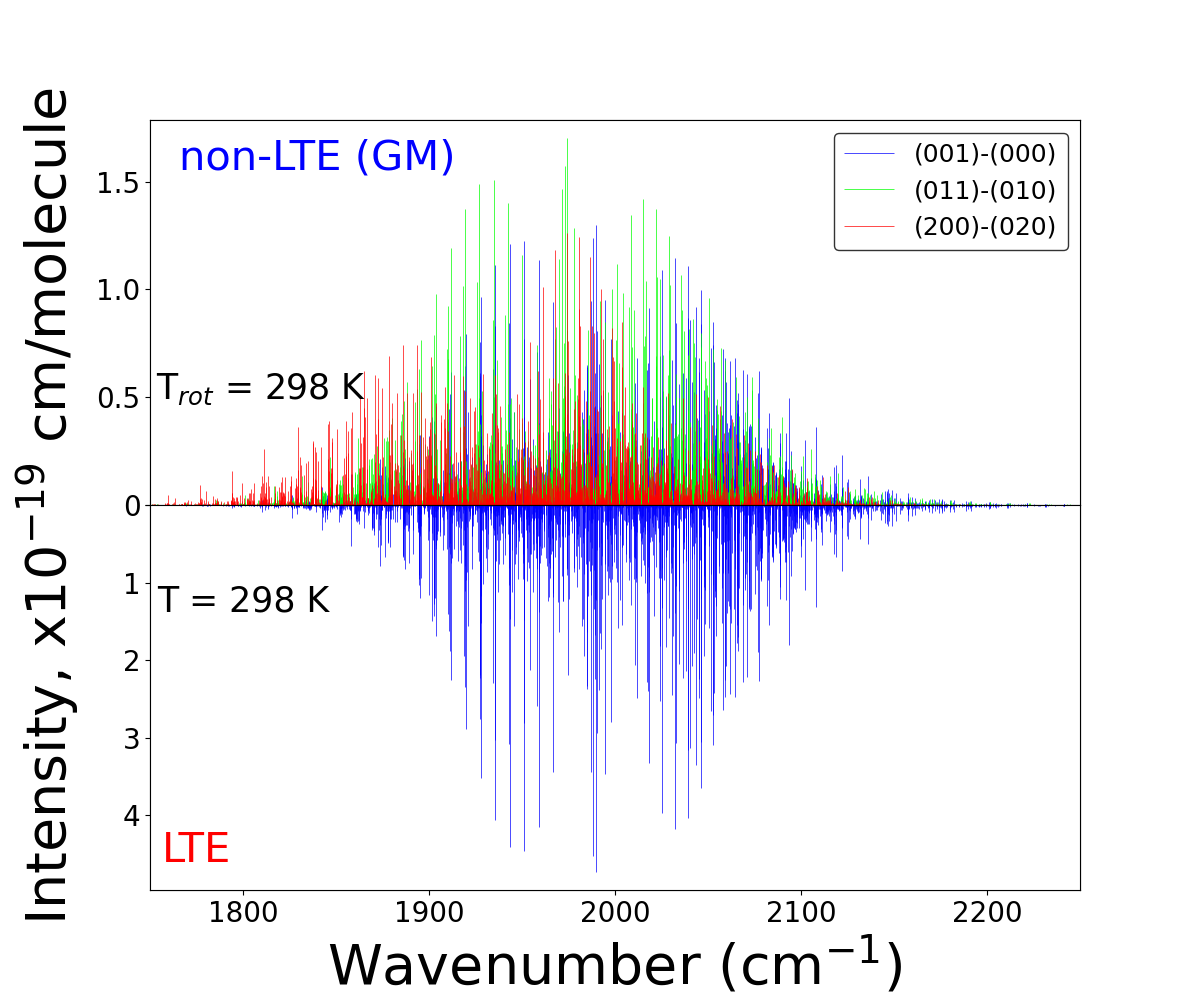}
\includegraphics[width=0.6\columnwidth]{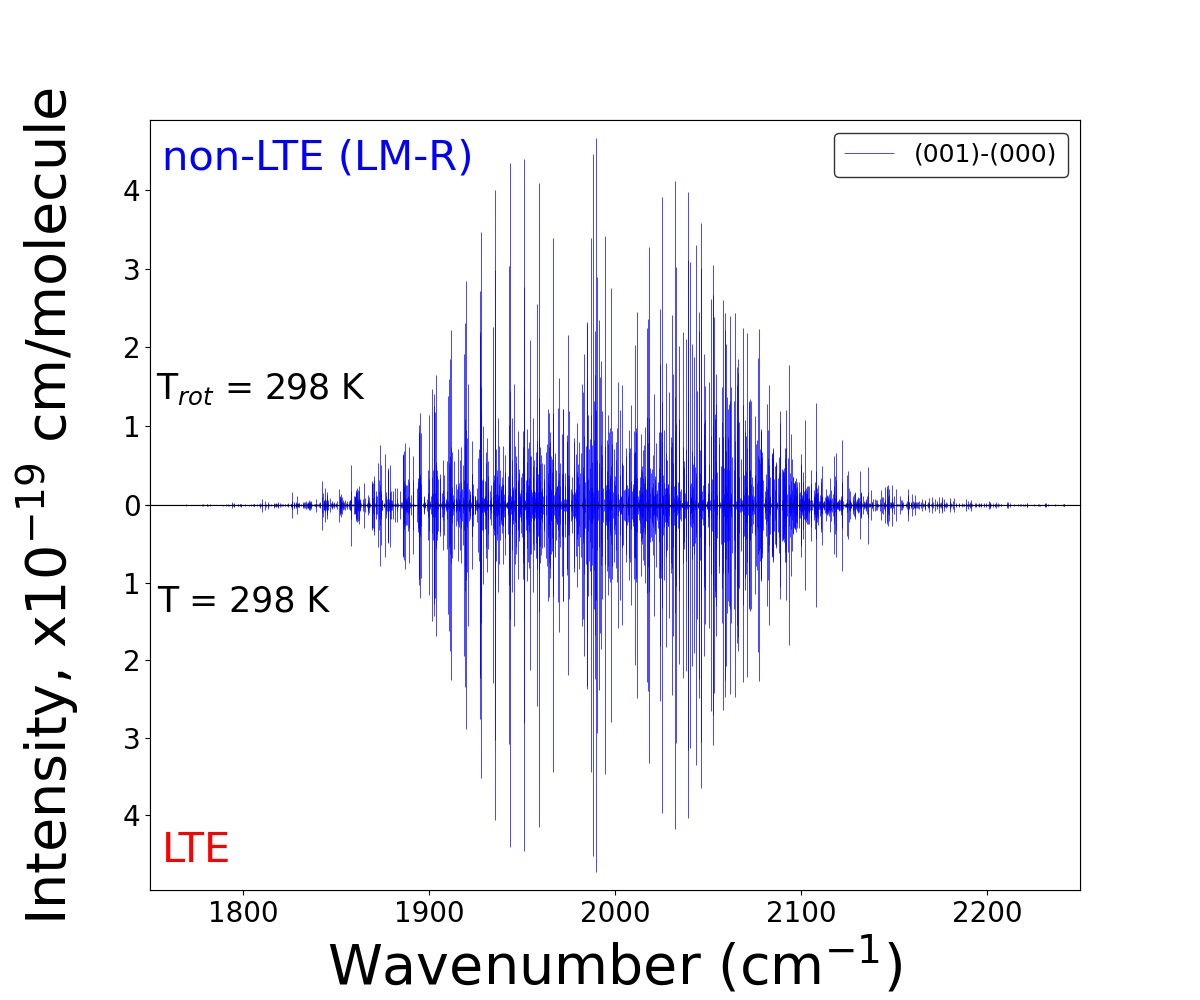}
\includegraphics[width=0.6\columnwidth]{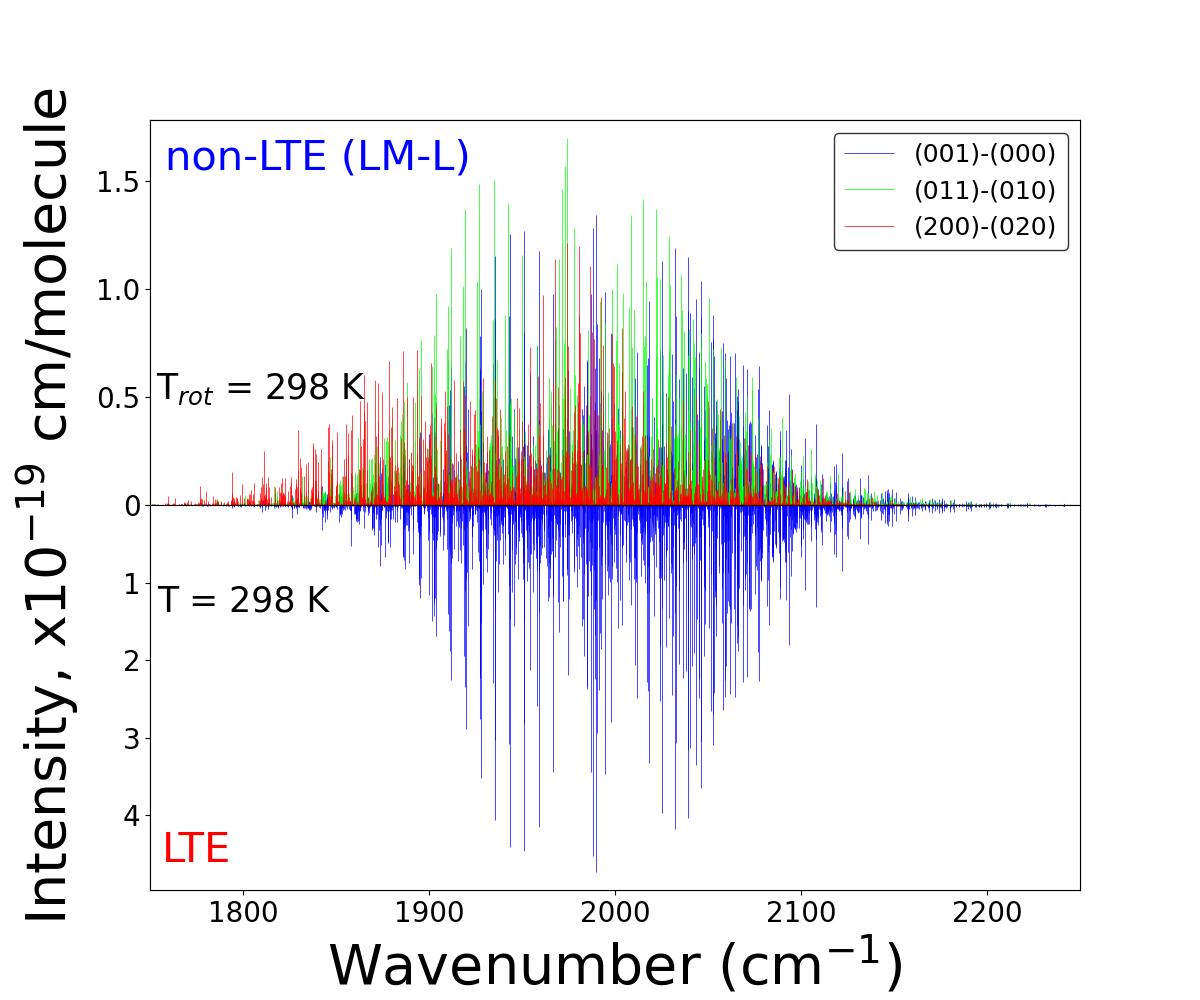}
\includegraphics[width=0.6\columnwidth]{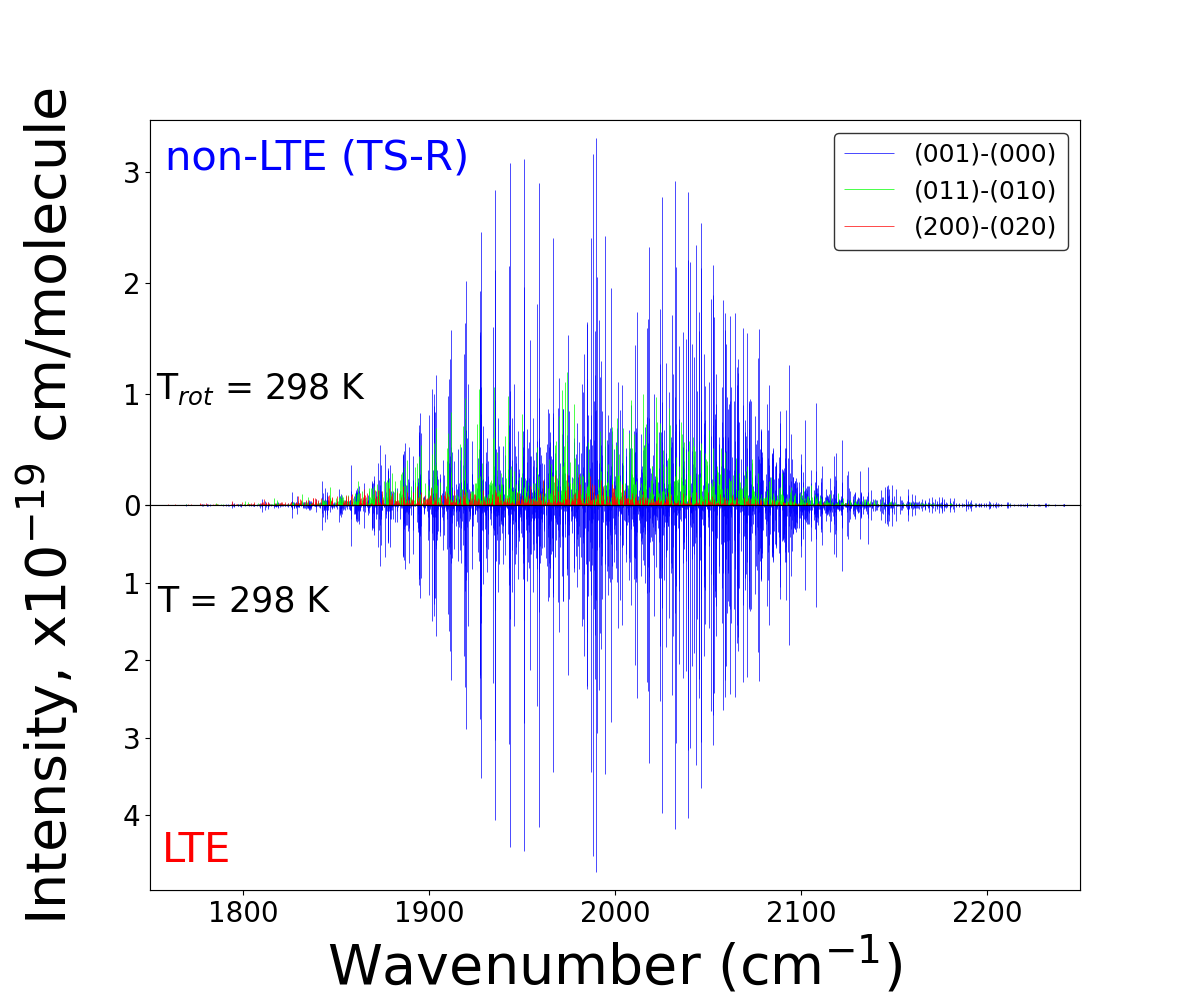}
\includegraphics[width=0.6\columnwidth]{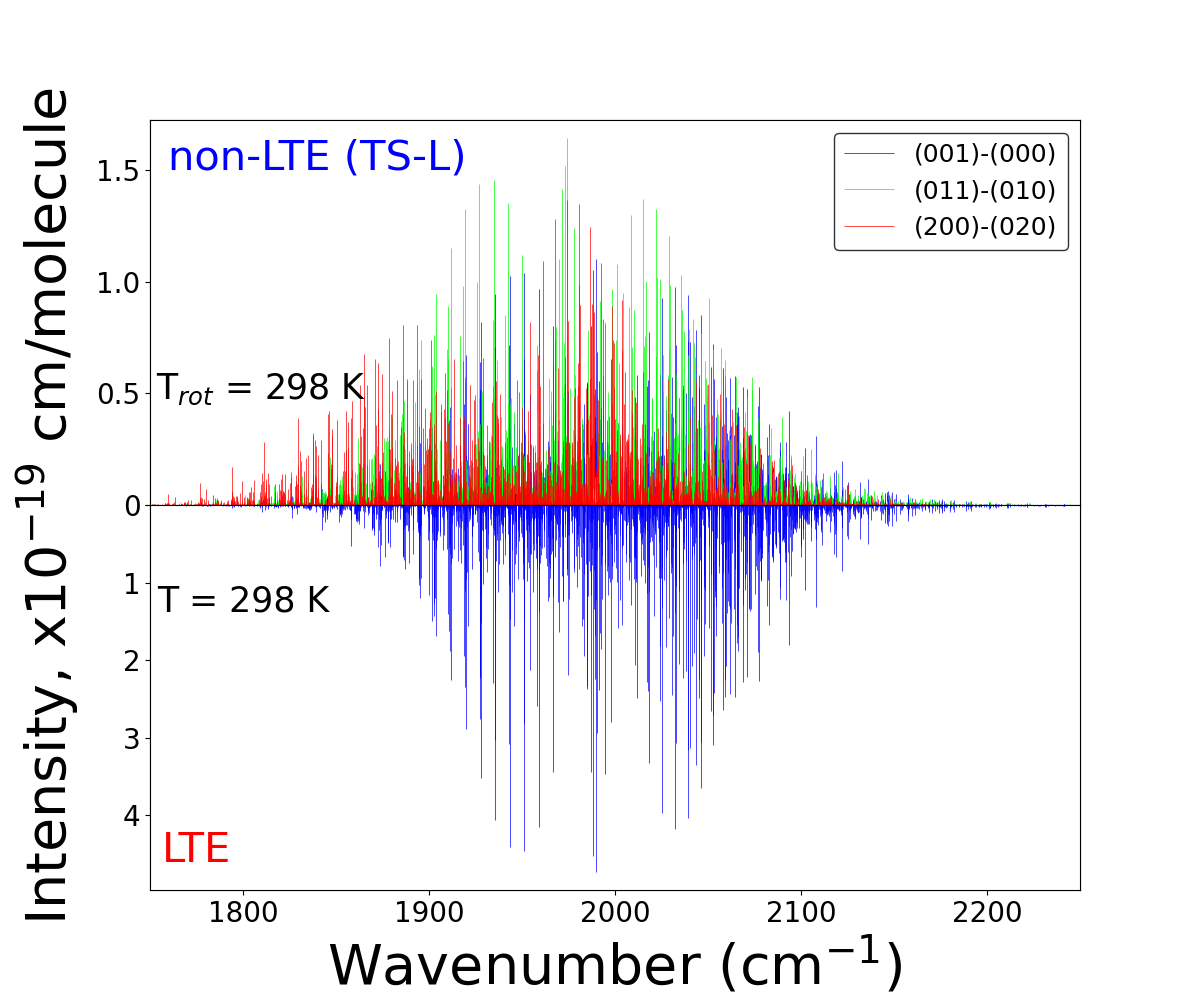}
\caption{Non-LTE spectra of SiH\2\ at $T_{\rm rot}$ = 296~K in the 2000~\icm\ (5~$\mu$m) region corresponding to five vibrational populations  GM, TS-L, TS-R, LM-L and LM-R (upper displays of each figure) and compared to the same spectra simulated using the LTE population at $T=296$~K. The non-LTE populations were obtained using the 1D approach (see text). Only the strongest bands are shown. The CATS line list was used }
\label{fig:filter-1d-2000}
\end{figure*}

Figures \ref{fig:filter-1d-1000} and \ref{fig:filter-1d-2000} show the 1000~\icm\ and 2000~\icm (10~$\mu$m and 5~$\mu$m) bands for the SiH\2\ absorption spectrum, respectively, simulated using the non-LTE densities from Fig.~\ref{fig:2d-populations-python} for all five cases considered and compared to the LTE scenario assuming the (rotational) temperature of $T$ = 296~K and using the CATS line list. The strongest bands are indicated using different colours.
Figure~\ref{fig:filter-1d-1000} focuses on the 1000~\icm\ band. Most of the non-LTE spectra contain bending hot bands (020)--(010) and (020)--(010), which are stronger than the fundamental band (000)--(000).  It can be seen that the $P$ and $R$ branches of the non-LTE spectra are shifted to lower wavenumbers in the GM, LM-L and TS-L spectra.  In the TS-R and LM-R spectra the bands are not shifted, with the LM-R spectrum having only the fundamental (010)--(000) band visible.

The plots in Fig. \ref{fig:filter-1d-2000} show the 2000~\icm\ band in the region of the polyad (100)/(020)/(001) for the five fragments, with the strongest fundamental band (001)--(000). The non-LTE intensities of the hot bands (011)--(010), (200)--(020) are found to be comparable to the intensities of the (001)--(000) band.  The $Q$ branch is clearly shifted for the GM, TS-L and LM-L molecules.  The band is less shifted for the TS-R and LM-R fragments, but owing to the increased similarity between the fragment and molecular structures with TS-R and LM-R this is to be expected. Only the main polyad system (100)/(020)/(001) is visible for the TS-R spectrum (indicated as (001)--(000) in Fig.~\ref{fig:filter-1d-2000}).

With the equilibrium structures of the TS-R and LM-R fragments being  similar to the equilibrium structure of \sihh, their non-LTE spectra are expected to be a similar spectrum to LTE.  Indeed, for the 1D harmonic approach   their $P$, $Q$ and $R$ branches maintain the expected LTE intensities for both the 1000~\icm\ and 2000~\icm\ bands.

\subsection{The 3D approach for vibrational populations using an accurate variational method}
\label{sec:3d-approach}

In a full 3D approach, the ground state wavefunction $\ket{0,0,0({\rm fragment})}$ represents an SiH\2\ fragment of an 18D ground state vibrational  wavefunction of Si\2H\6:
$$
\phi_{{\rm Si}_2{\rm H}_6} = \ket{0,0,0({\rm SiH}_2)} \ket{0,0,0,0,0,0,0,0,0,0,0,0,0,0,0({\rm SiH}_4)}.
$$
Here we assume the approximation that the corresponding  three modes (Si--H$_1$, Si--H$_2$ and $\angle$H$_1$SiH$_2$) are independent from the rest of the molecule so that all other modes, not relevant for the gas phase SiH\2, can be eliminated (integrated out),  including the reaction coordinate  and vibrational modes of  SiH\4. This is in line with the assumptions used previously by \citet{75BaFrxx} and \citet{74Berryx}. Apart from this approximation we will treat the SiH\2\ fragment as accurate as possible. The corresponding wavefunction
$\ket{0,0,0({\rm fragment})}$ is obtained by solving a 3D vibrational Schr\"{o}dinger equation for these three degrees of freedom with a realistic PES obtained using a high level of \ai\ theory (the same as above, VTZ/CCSD(T)-F12b with MOLPRO).

The vibrational populations are then modelled using the Franck-Condon integrals as follows:
\begin{equation}
\label{e:N=<0|v>^2:3D}
N_{\varv_1,\varv_2,\varv_3} = |\langle {0,0,0} ({\rm fragment})  \ket{\varv_1,\varv_2,\varv_3({\rm g.ph.})}|^2,
\end{equation}
with the sum of all populations over all states totalling 1.
In this equation, $\ket{\varv_1,\varv_2,\varv_3({\rm g.ph.})}$ is an accurate vibrational wavefunction of a gas phase SiH\2\ molecule, obtained by solving the vibrational Schr\"{o}dinger equation with an accurate PES. We  use the \trove\ variational  program and the refined PES of SiH\2\ by \citet{jt779}  to generate  $\ket{\varv_1,\varv_2,\varv_3({\rm g.ph.})}$  for all vibrational excitations required.  For the details on the \trove\ calculations  see below and also \citet{jt779}

In order to simplify the 3D integration in Eq.~\eqref{e:N=<0|v>^2:3D}, the variational  wavefunction  $\ket{0,0,0({\rm fragment})}$  is obtained using the same vibrational basis set as the variational solution of the gas phase SiH\2. By taking advantage of the  compatibility of the orthogonality of the basis sets, the Franck-Condon factors are then given by as a sum of products
\begin{equation}
\label{e:FC=CxC}
\langle {0,0,0} ({\rm fragment})  \ket{\varv_1,\varv_2,\varv_3({\rm g.ph.}}) = \sum_{i_1,i_2,i_3} C^*_{i_1,i_2,i_3}({\rm fragment}) \, C_{i_1,i_2,i_3}({\rm g.ph.})
\end{equation}
of the corresponding eigen-coefficients  $C_{i_1,i_2,i_3}({\rm fragment})$ and $C_{i_1,i_2,i_3}({\rm g.ph.})$, obtained variationally in independent calculations using a new implementation in \trove.

\trove\ uses optimized non-standard vibrational basis sets, generated numerically  by solving 1D Schr\"{o}dinger equations for realistic 1D potentials.\citep{TROVE} This procedure allows producing compact basis functions optimized for a specific problem. In our case, the  PESs of the  corresponding five fragments and of the gas phase SiH\2\ are different and therefore the generated  basis sets would be different and even not orthogonal. We therefore implemented a feature in TROVE allowing to read and use externally generated basis functions. Of course all relevant  calculation setups must be compatible, including the numerical grids used for the stretching and bending modes and  their sizes. Using foreign basis sets certainly degrades their quality. However, since we are only interested in fragments' ground state wavefunctions, this degradation can be mitigated by including enough basis functions. Our typical 1D basis sets contain 12 -- 24 functions (see details below), which should be more than enough to obtain a converged ground state solution even with non-optimized basis sets.

\subsubsection{PESs of SiH\2\ fragments}

For our new 3D populations corresponding to dissociations from the fragments,  five PESs were generated as follows. We assume that PESs of a dissociating  Si\2H\6\ molecule  can be approximated as a sum of two independent fragments:
$$
V_{{\rm Si}_2{\rm H}_6} = V_{{\rm SiH}_2}(r_1,r_2,\alpha) + V_{\rm rest},
$$
where the individual stretching and bending modes of  \sihh\ fragments  are  fully separable:
\begin{equation} \label{eq:pes-representation}
V_{{\rm SiH}_2}(r_1,r_2,\alpha) = f_{\rm str}(r_1) + f_{\rm str}(r_2) + f_{\rm bnd}(\alpha).
\end{equation}
 The stretching part of the potential is given by a Morse-like expansion
\begin{equation} \label{eq:f-stretch}
f_{\rm str}(r) = \sum_{i} f_i \left[ 1-e^{-a\Delta (r-r_{\rm e})}\right]^i,
\end{equation}
while the bending part is a Taylor-type expansion in terms of the displacement from the corresponding equilibrium value:
\begin{equation} \label{eq:f-bend}
f_{\rm bnd}(\alpha) = \sum_{i} g_i (\cos\alpha-\cos\alpha_e)^i .
\end{equation}

The expansion constants $f_i$ and $g_i$ representing fragments' potential energies  $V_{{\rm SiH}_2}(r_1,r_2,\alpha)$  were obtained by fitting  Eqs.~(\ref{eq:f-stretch},\ref{eq:f-bend})
to the   \ai\ data computed as 1D slices on the global surfaces for the five fragments from \disil\   (\textit{d}GM, \textit{d}LM and \textit{d}TS)   using VTZ/CCSD(T)-F12b consisting of 24 bending and 34 stretching geometries, distributed around the corresponding equilibria.

\begin{table*}[h]
\caption{\label{t:f-and-g} The \ai\ potential parameters defining the 1D potential of different structures  used in \trove\ calculations. The units are \icm,  \AA\ and radians, unless specified}
\begin{tabular}{c d{8.5} d{8.5} d{8.5} d{8.5} d{8.5}}
\hline
Parameter & \textrm{GM} & \textrm{TS-L} & \textrm{TS-R} & \textrm{LM-L} & \textrm{LM- R}\\
\hline
$r_{\rm e}$, \AA & 1.48218 & 1.47971 & 1.49888 & 1.47982 & 1.51611 \\
$\alpha_{\rm e}$, deg & 108.701 & 108.938 & 103.268 & 108.264 & 92.287 \\
$a$, \AA$^{-1}$  & 1.29065 & 1.29110 & 1.22364 & 1.27911 & 1.27050 \\
$g_1$		&	0	&	0	&	0	&	0	&	-8132.4642	\\
$g_2$		& 80964.377	&	90039.635	&	101613.84	&	93608.413	&	168296.56	\\
$g_3$		&	-117331.83	&	-126539.98	&	-159054.10	&	-129420.43	&	-299843.15	\\
$g_4$		&	139212.98	&	168090.27	&	182459.45	&	305247.60	&	324286.87	\\
$f_1$		&	0	&	0	&	0	&	0	&	-793.07685\\
$f_2$		&	43617.985	&	43796.459	&	45462.423	&	22217.055	&	19382.230	\\
$f_3$		&	-6594.4391	&	-6937.7586	&	-8917.729	&	-3824.8354	&	-2298.2220	\\
$f_4$		&	2809.6655	&	2714.6696	&	3750.0966	&	1407.7267	&	2233.5205	\\
\hline
\end{tabular}
\end{table*}

\begin{figure*}[ht]
\centering
\includegraphics[width=0.5\columnwidth]{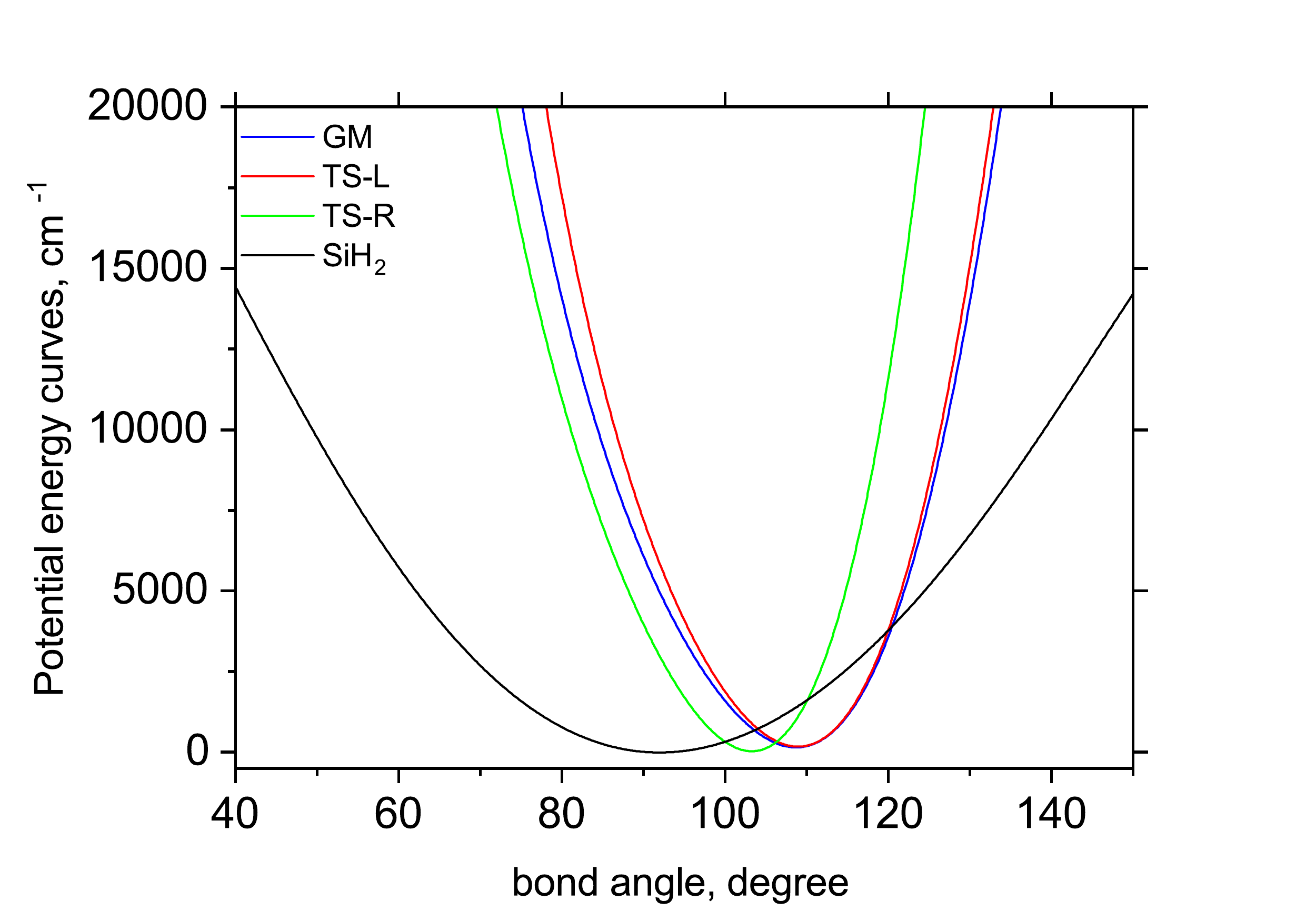}
\includegraphics[width=0.5\columnwidth]{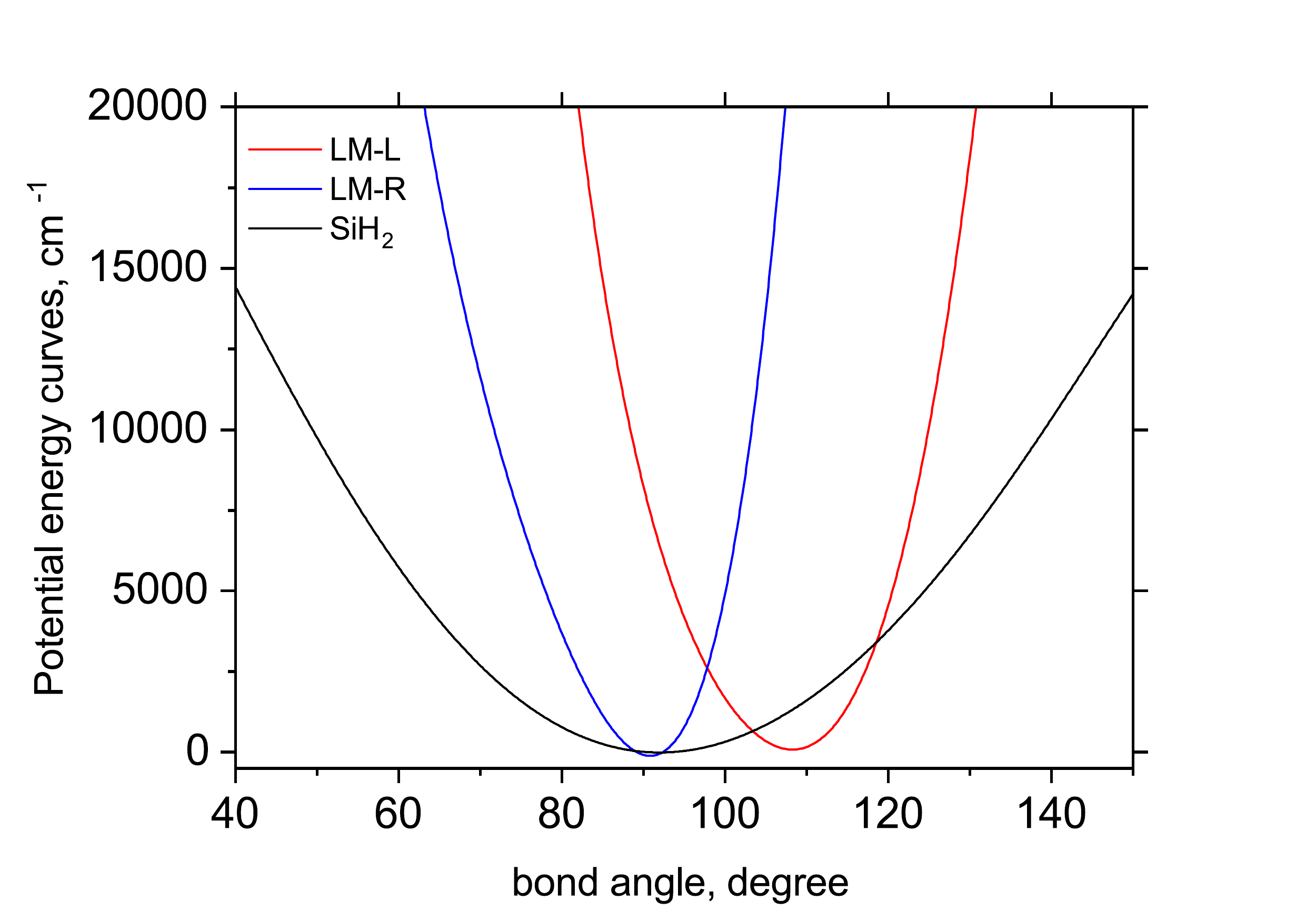}
\includegraphics[width=0.5\columnwidth]{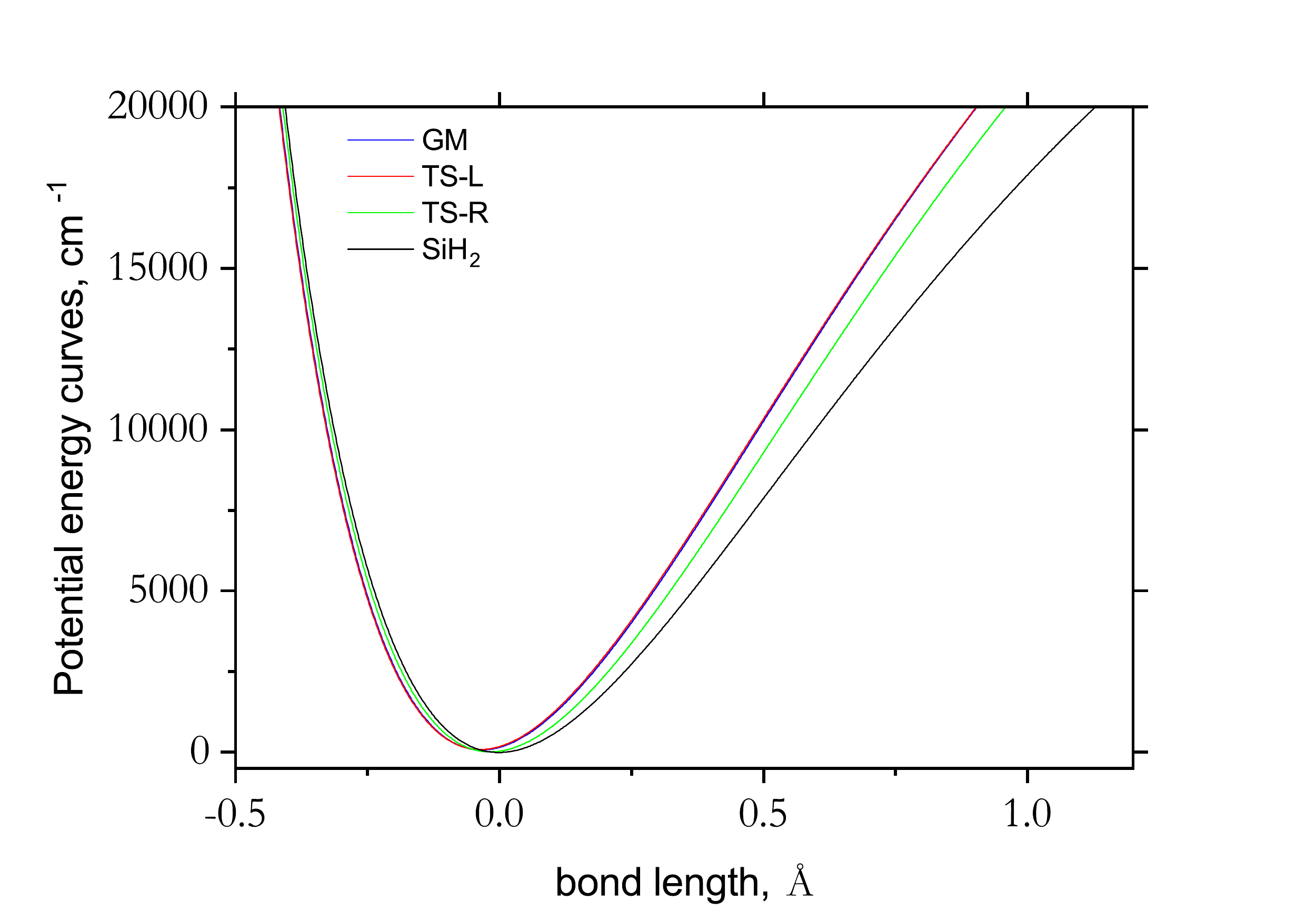}
\includegraphics[width=0.5\columnwidth]{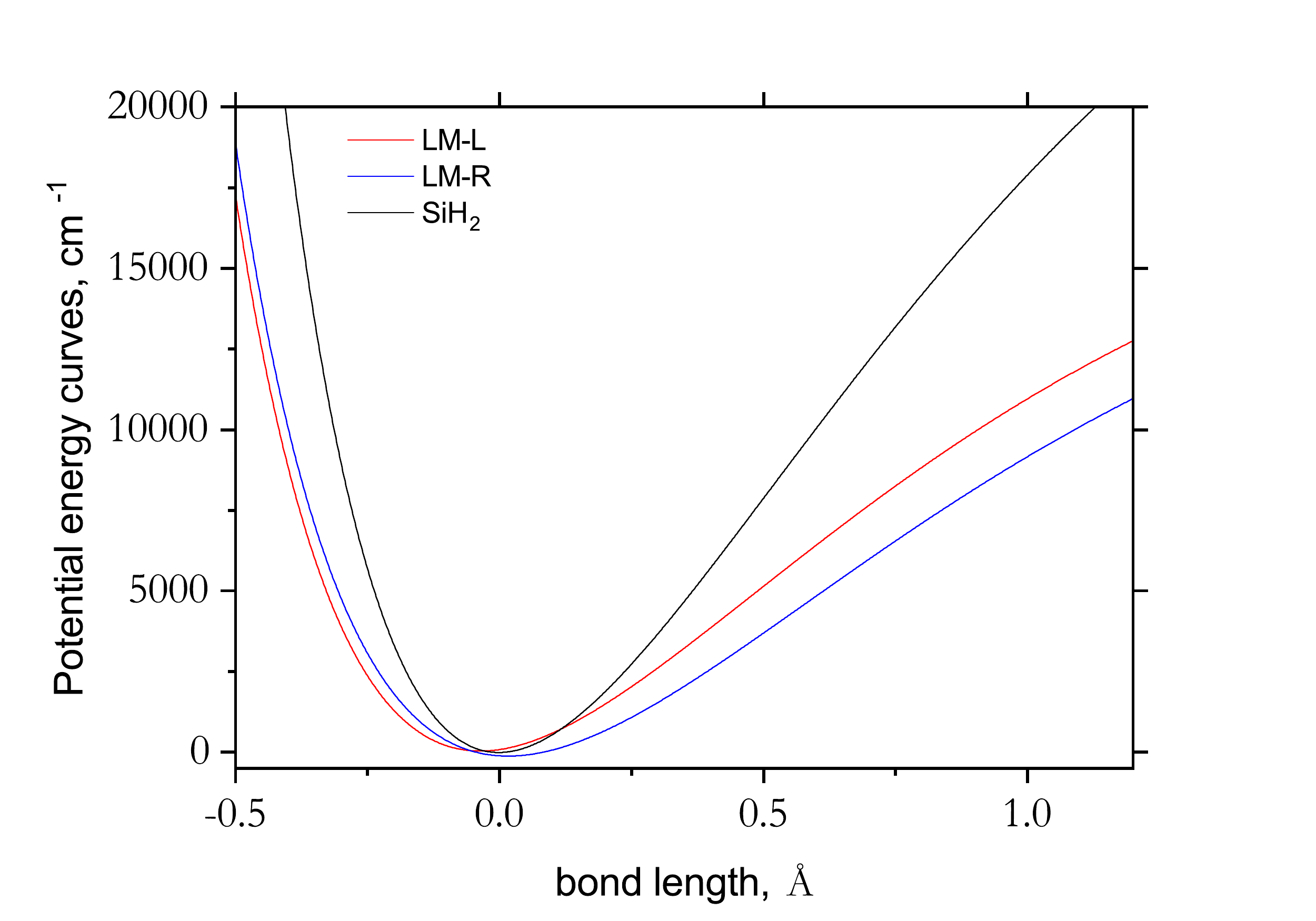}
\caption{ 1D potential energy cuts  representing different stretching Si--H ($x$-axis = bond length)  and bending ($x$-axis = bond angle) H--Si--H modes of different fragments of three Si\2H\6\ isomers, \textit{d}GM, \textit{d}LM and \textit{d}TS, compared to the corresponding cuts of the gas phase SiH\2\ species}
\label{f:1D:PES:Fragments}
\end{figure*}

Figure~\ref{f:1D:PES:Fragments} illustrates the \ai\ PESs of different fragments as 1D cuts for the stretching and bending modes compared to the corresponding cuts of the gas phase SiH\2\ molecule. The bending cuts have  especially different shapes with  shifts to larger equilibrium angles and much steeper PESs. The differences in the stretching cuts are less pronounced. These features are important for the non-LTE behaviour of the corresponding excited states populations, with the bending degree of freedom to have stronger non-LTE character than stretching.

\subsubsection{Vibrational calculations}

The vibrational wavefunctions of  SiH\2\ were computed using the variational nuclear motion program \trove\  with  the same setup as in  \citet{jt779}. Details of the \trove\  methodology are discussed extensively elsewhere.\cite{TROVE,15YaYuxx.method,17YuYaOv.methods,jt626}
Here, we give a brief outline of the  main calculation steps. The \trove\ kinetic energy operator is Taylor expanded up to sixth order  around the SiH$_2$ equilibrium geometry in terms of  linearized coordinates.\citep{98BuJe.method}
The  primitive basis set is constructed from  1D mode numerical basis functions using the Numerov-Cooley approach~\citep{24Numerov.method,61Cooley.method} by solving three 1D Schr\"{o}dinger equations, for each vibrational degree of freedom. The stretching basis functions are then improved by solving a 2D Schr\"{o}dinger equation for a reduced stretching Hamiltonian.  The resulting stretching eigenfunctions  are contracted, classified according with the \Cv{2}(M) symmetry group~\citep{98BuJe.method} using an optimized symmetrization procedure~\citep{17YuYaOv.methods} and combined with the bending primitive basis functions to form our final, symmetry-adapted 3D vibrational basis set.
The  basis set coverage is defined by the polyad number  cut-off
\begin{equation}
P = 2 (\varv_1 + \varv_3) + \varv_2 \le 24,
\end{equation}
where  $\varv_1$ and $\varv_3$ are the stretching and $\varv_2$ is the bending quantum numbers, with the maximal excitations 12, 12 and 24 respectively.

For the gas phase SiH\2\  calculations we employed the empirically refined PES by \citet{jt779} For the five SiH\2\ fragments only the vibrational ground state wavefunctions $\ket{0,0,0}$ were computed using the same setup and utilizing the basis functions from the g.ph. calculations as described above, but for the fragments' \ai\ PESs from Eq.~\eqref{eq:pes-representation}.

\subsubsection{Vibrational populations of SiH\2}

The 3D vibrational populations $N_{\varv_1,\varv_2,\varv_3}$  in Eq.~\eqref{e:N=<0|v>^2:3D} were computed using the corresponding TROVE eigenfunctions via Eq.~\eqref{e:FC=CxC}.

\begin{figure}[ht]
\centering
\includegraphics[width=0.7\columnwidth]{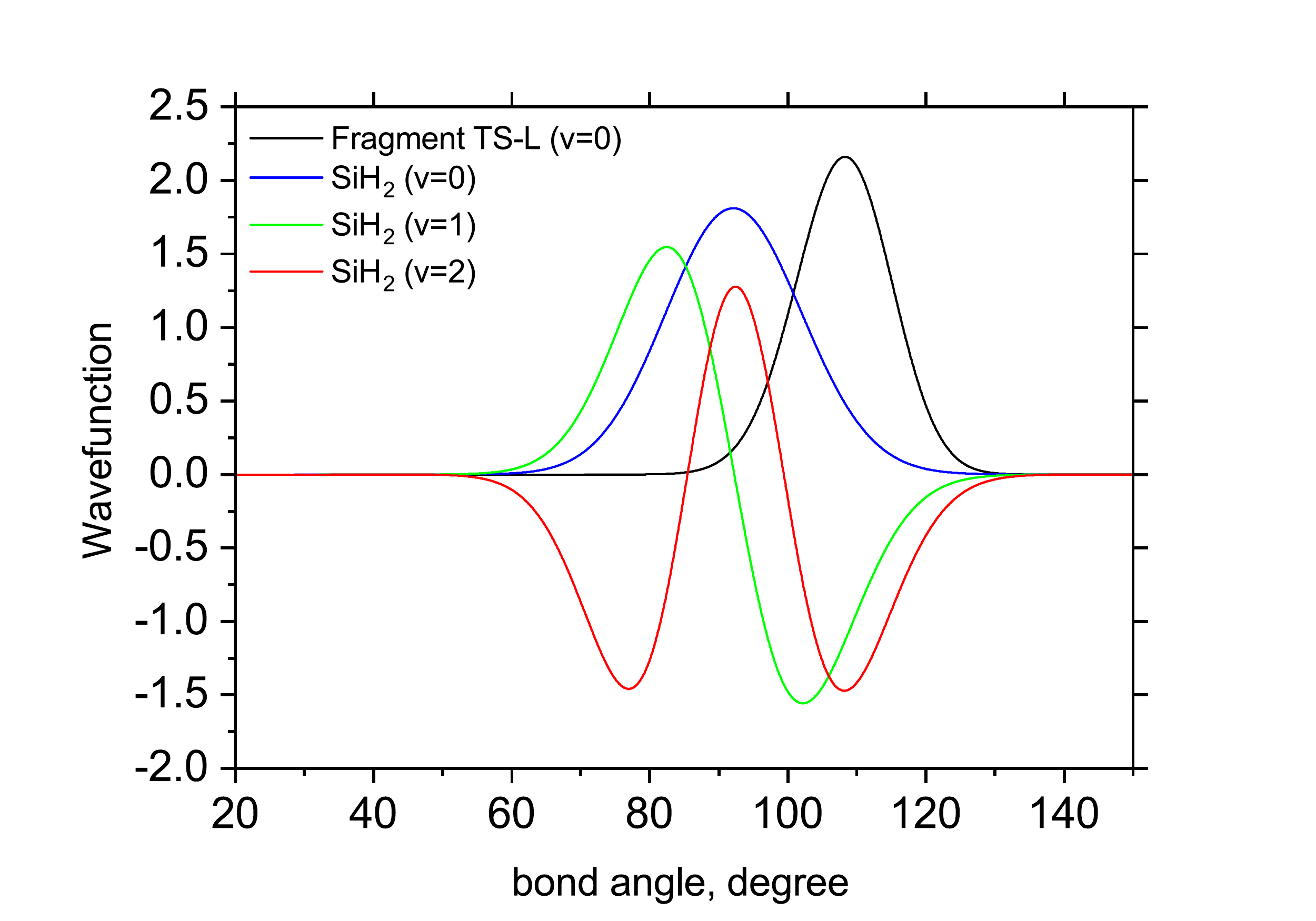}
\caption{ 3D Bending primitive wavefunctions of SiH\2\ compared to the ground state 3D bending  wavefunctions for Fragment TS-L. Calculated with \TROVE}
\label{f:bnd:wave:F2}
\end{figure}

\begin{figure}[ht]
\centering
\includegraphics[width=0.7\columnwidth]{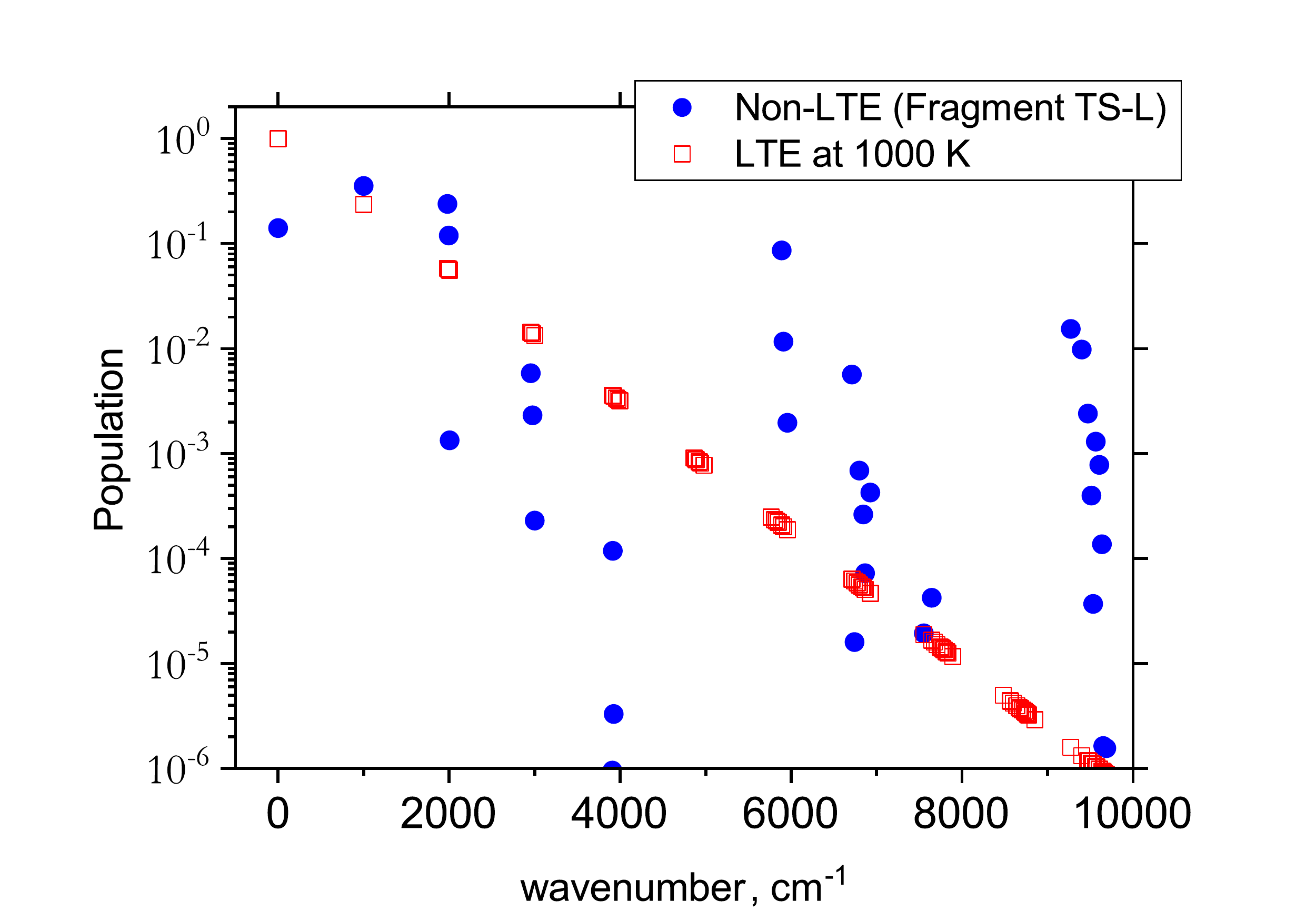}
\caption{The 3D non-LTE vibrational populations of SiH\2\ for the dissociation from TS-L (blue circles) as a function of the vibrational term values compared to Boltzmann distribution of  SiH\2\ vibrational states at $T$ = 1000~K}
\label{f:3D:populations}
\end{figure}

The 3D results generally agree with the 1D. As an example, Fig.~\ref{f:bnd:wave:F2} shows the TS-L case: The bending primitive wavefunctions of the gas phase SiH\2\ are compared to the $\varv=0$ bending wavefunction generated for the PES of TS-L. Figure~\ref{f:3D:populations} shows the corresponding  non-LTE vibrational populations as a function of the corresponding energies for the same TS-L scenario of the 3D vibrational populations. These populations are very different from the Boltzmann distribution, also shown on this figure for the  SiH\2\ vibrational states at $T$ = 1000~K, which exhibits an exponential decay  with  $\varv=0$ at its maximum. It nicely demonstrates  that it would not be possible to associate a single  vibrational temperature for this bell-shaped distribution.

\subsection{Non-LTE intensity simulations}
\label{sec:non-LTE-simulations}
A non-LTE absorption line intensity $I_{\rm fi}$ (cm/molecule) can be calculated as
\begin{equation}
\label{e:int}
 I({\rm f} \gets {\rm i}) = \frac{g_f^{\rm tot} A_{\rm fi}}{8 \pi c \tilde\nu_{\rm fi}^2}  \frac{N_{J,k,\varv} \left( 1-e^{-c_2\tilde{\nu}_{\rm fi}/T} \right)}{Q_{\rm nLTE}(T)},
\end{equation}
where $A_{\rm fi}$ is the Einstein-A coefficient ($s^{-1}$), $\tilde{\nu}_{\rm fi}$ is the transition wavenumber (\cm), $Q_{\rm nLTE}(T)$ is the non-LTE partition function defined in Eq.~\eqref{e:pf}.

In order to simulate  absorption spectra of the gas phase SiH\2\  assuming  a non-LTE vibrational populations $N_{\varv_1,\varv_2,\varv_3}$, we use the line positions and Einstein A coefficients from the ExoMol CATS line list \cite{jt779} employing the \xc\ program.\citep{ExoCross} The ExoMol line lists are formatted as two files, a States file and a Transition file.  It is described extensively elsewhere \cite{jt810} and in this paper we shall only discuss how the ExoMol format has been adapted for use in non-LTE situations.

To adapt the CATS  States file \citep{jt779} for non-LTE applications an additional `density' column $N_{\varv_1,\varv_2,\varv_3}$  was added  as a final column, see an extract from the States file in Table~\ref{tab:states-file}.  This column contains the weightings to the transition probabilities as  populations of the vibrational levels occupied by the gas phase \sihh.  The `density' column  is specific for the calculation of line lists for non-LTE molecules is not routinely included into the ExoMol States files.

This column is read by \xc\ and used to give the population  weighting  to each line intensity. Evaluation of the non-LTE population $ N_{J,k,\varv}(T)$ given for each ro-vibrational state as in Eq.~\eqref{e:population} is based on the knowledge of  the corresponding vibrational state $(\varv_1,\varv_2,\varv_3)$ as well as the rotational energy contribution $\tilde{E}_{J,k}^{\rm rot}$. Therefore for this approach to work it is mandatory for all ro-vibrational states  to be vibrationally assigned in order to be able to subtract the vibrational contribution $\tilde{E}_{\varv}^{\rm vib}$ from the total energy  according with Eq.~(\ref{e:E=Evib+Erot}). All vibrational quantum numbers ($\varv_1,\varv_2,\varv_3$) are not required, only a vibrational index $\varv$ indicating the vibrational state in question. In our model, all asymmetric vibrational states ($B_2$) are not populated due to the zero overlap with the ground state of the $A_1$ symmetry in Eq.~\eqref{e:N=<0|v>^2:3D}, as part of the completely vertical Franck-Condon approximation.

\begin{table*}
\caption{\label{tab:states-file} Extract from the modified \texttt{.states} file of the GM 1D non-LTE line list}
{\tt\footnotesize
\begin{tabular}{rrrrrrrrrrrrrrrrrc}
\hline
\multicolumn{1}{c}{$i$} & \multicolumn{1}{c}{$\tilde{E}_i$} & $g_i$ & $J$ & $\Gamma_{\textrm{tot}}$ & $\varv_1$ & $\varv_2$ & $\varv_3$ & $\Gamma_{\textrm{vib}}$ & $K_a$ & $K_c$ & $\Gamma_{\textrm{rot}}$ & $C_i$  & $n_1$ & $n_2$ & $n_3$ & $i_{\rm vib}$ & $N_{\varv}^{\rm vib}$\\
\hline
1527 & 19978.952397 & 3 & 1 & A2 &  2 &  8 & 5 & B2 & 0 & 1 & B1 &  0.98 & 2 &  5 &  8 & 624 &  2.01226E-14 \\
1528 & 19982.538080 & 3 & 1 & A2 &  0 & 20 & 1 & B2 & 0 & 1 & B1 &  0.98 & 0 &  1 & 20 & 625 &  9.45465E-15 \\
1529 & 19986.504561 & 3 & 1 & A2 & 10 &  2 & 1 & B2 & 0 & 1 & B1 & -0.99 & 0 & 11 &  2 & 626 &  4.24271E-16 \\
1530 & 19988.143253 & 3 & 1 & A2 & 11 &  2 & 0 & A1 & 1 & 1 & A2 &  0.99 & 0 & 11 &  2 & 627 &  4.24271E-16 \\
1531 &    10.721260 & 9 & 1 & B1 &  0 &  0 & 0 & A1 & 0 & 1 & B1 &  1.00 & 0 &  0 &  0 &   1 &  2.53409E-01 \\
1532 &  1009.393331 & 9 & 1 & B1 &  0 &  1 & 0 & A1 & 0 & 1 & B1 & -1.00 & 0 &  0 &  1 &   2 &  3.11385E-01 \\
1533 &  1989.081019 & 9 & 1 & B1 &  0 &  2 & 0 & A1 & 0 & 1 & B1 &  1.00 & 0 &  0 &  2 &   3 &  2.10395E-01 \\
1534 &  2004.596674 & 9 & 1 & B1 &  0 &  0 & 1 & B2 & 1 & 1 & A2 & -1.00 & 0 &  1 &  0 &   4 &  8.56926E-03 \\
1535 &  2016.232068 & 9 & 1 & B1 &  1 &  0 & 0 & A1 & 0 & 1 & B1 &  1.00 & 1 &  0 &  0 &   5 &  8.56926E-03 \\
1536 &  2962.467155 & 9 & 1 & B1 &  0 &  3 & 0 & A1 & 0 & 1 & B1 & -1.00 & 0 &  0 &  3 &   6 &  1.01925E-01 \\

\hline
\end{tabular}
}
{\flushleft \footnotesize
$i$:   State counting number; $\tilde{E}_i$: State energy in \cm; $g_i$: State degeneracy; $J$: Total angular momentum quantum number; $\Gamma_{\textrm{tot}}$: Overall symmetry of state in \Cv{2}(M); $\varv_1$--$\varv_3$:  Vibrational (normal mode) quantum numbers; $\Gamma_{\textrm{vib}}$: Vibrational symmetry in \Cv{2}(M); $K_a$:  Asymmetric top quantum number; $K_c$:  Asymmetric top quantum number; $\Gamma_{\textrm{rot}}$: Rotational symmetry in \Cv{2}(M); $C_{i}$: Largest coefficient  used in the \TROVE\ assignment; $n_1$--$n_3$:  Vibrational (TROVE) quantum numbers; $i_{\rm vib}$: Vibrational state counting number; $N_{\varv}^{\rm vib}$ Population density for each $i$ based on $\varv_1$--$\varv_3$. This column is not produced by the ExoMol format. It must be calculated separately.
}
\end{table*}

\subsubsection{Using the non-LTE populations from the 3D approach}

\begin{figure}[ht]
\centering
\includegraphics[width=0.95\columnwidth]{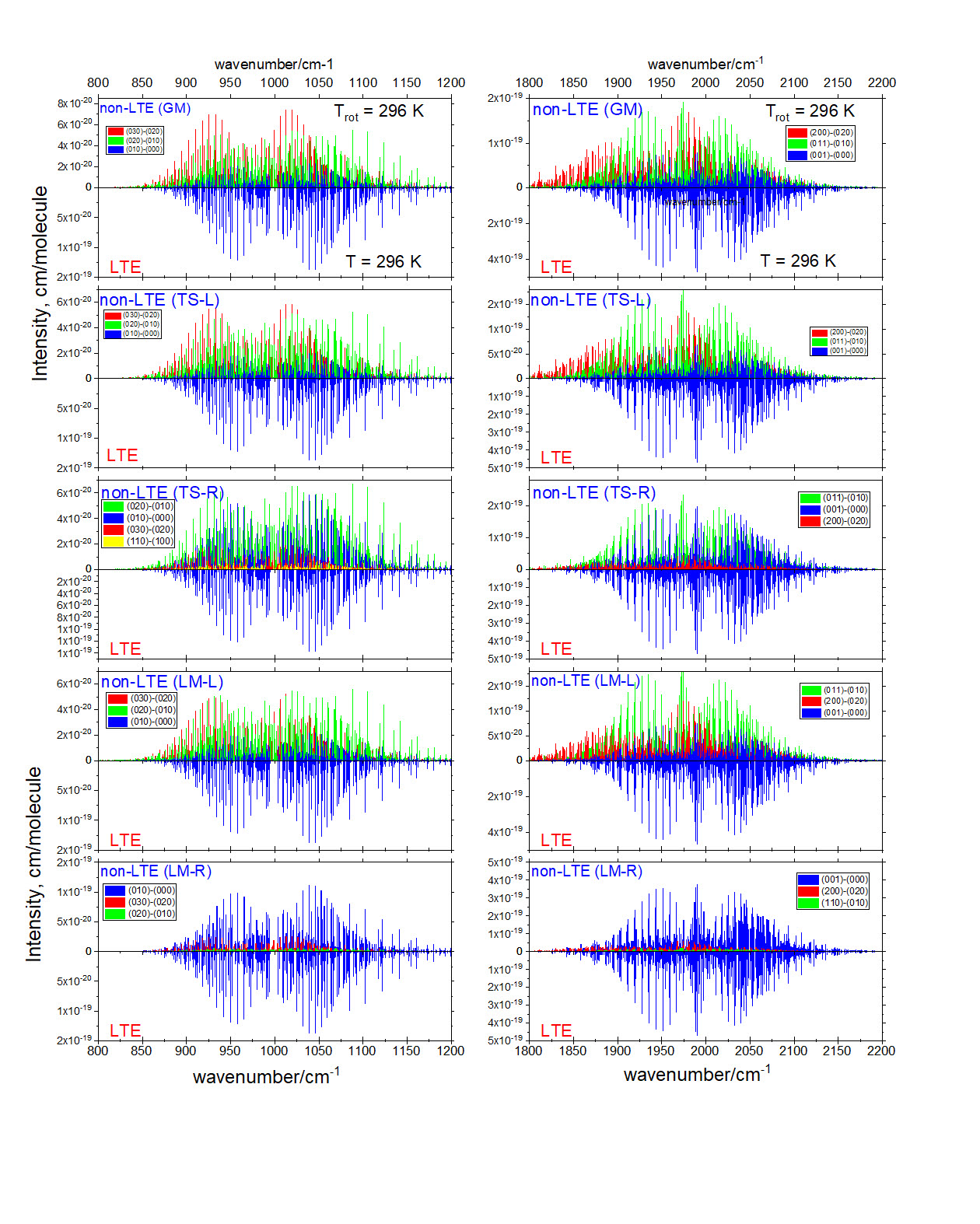}
\caption{Non-LTE spectra of SiH\2\ at $T_{\rm rot}$ = 296~K corresponding to five  vibrational populations  GM, TS-L, TS-R, LM-L and LM-R (upper displays of each figure) and compared to the same spectra simulated using the LTE population at $T=296$~K. The non-LTE populations were obtained using the 3D TROVE approach. Only three or four strongest  bands are shown. The CATS line list was used  }
\label{fig:non-LTE-3D:stick:10}
\end{figure}

Figure~\ref{fig:non-LTE-3D:stick:10} shows  non-LTE spectra of SiH\2\ in the two main spectroscopic regions, 1000 and 2000~\icm\ for all  five fragments considered (GM, TS-L, TS-R, LM-L and LM-R). The corresponding vibrational non-LTE populations were  generated with the 3D TROVE approach. The rotational populations  assume the  Boltzmann distribution with the rotational temperature of  $T_{\rm rot} = 296$~K.  These non-LTE spectra are compared to the corresponding LTE spectra of SiH\2\ at $T= $ 296~K, which  comprise mainly two fundamental bands, $(0,1,0)\leftarrow (0,0,0)$ (1000~\icm region) and $(0,0,1)\leftarrow (0,0,0)$ (2000~\icm region), with the hot bands suppressed due to the relatively low temperature. The non-LTE spectra are dominated by the hot bands  $(0,2,0)\leftarrow (0,1,0)$, $(0,3,0)\leftarrow (0,2,0)$ (1000~\icm\ band), $(1,1,0)\leftarrow (0,1,0)$ and  $(1,1,0)\leftarrow (0,2,0)$ (2000~\icm\ band). The centers of the hot bands are systematically red shifted compared to the fundamental band centres and serve as distinct signatures of the non-LTE effects. The $Q$-branch of $(011)\leftarrow  (020)$ is especially distinct compared to the $Q$-branch of the LTE fundamental band in the 2000~\icm\ region, with the difference of about 15~\icm.  Only the LM-R spectra of non-LTE are very similar to the LTE spectra. This is not surprising considering that the equilibrium values of $r_{\rm SiH}$ and $\alpha_{\angle {\rm H-Si-H}}$ of the LM-R structure are  very similar to the corresponding equilibrium parameters of the gas phase SiH\2.

The  stark differences of the non-LTE spectra offer an ability for experiment to distinguish between \sihh\ molecules produced from fragmenting \disil, and even to indicate dissociation channels involved.

\subsection{$A$--$X$ spectrum}
\label{sec:A-X-spectrum}

The visible electronic band $\tilde{A}$--$\tilde{X}$ of SiH\2\ has often been used to study different reactions involving  leading to silylene.\citep{98EsCaxx.SiH2} Here we used the program RENNER \citep{95JeBrKr.method} to simulate a non-LTE electronic  spectrum of SiH\2\ with the spectroscopic model by \citet{04YuBuKr.SiH2} for the \ab\ -- \xa\ system. The model includes two empirically adjusted PESs, for the $A$ and $X$ states, respectively, and an \ai\ (MRCI) $\tilde{A}$--$\tilde{X}$ transition dipole moment surface (TDMS).

In the RENNER calculations,  the size of the basis set originally used in \citet{04YuBuKr.SiH2} was reduced in order to be able to increase the rotational excitations. The main purpose of this exercise is to show a qualitative impact of the non-LTE populations on the spectral shape of the electronic band and not so much the quality of the line positions, and therefore a smaller basis set is justified. We used 18 and 12 bending basis functions for the $\tilde{X}$ and $\tilde{A}$ electronic
states, respectively, for every $|k|$ block (where $|k|$ $\leq$ $J$). The \xa\ electronic state basis set included  $N_A=$12  stretching functions
of the $A_1$ symmetry and  $N_B=$10  stretching functions
of the $B_2$ symmetry. For the \ab\ state $N_A=$10 and $N_B=$8 stretching functions were used.  These stretching functions were constructed from the Morse oscillator
functions $\vert n_{1}\rangle$  $\vert n_{3}\rangle $ with $n_1+n_3\le$ $N_{\rm stretch}=$12.

A rovibronic line list for the \ab\ -- \xa\ of SiH\2\ was generated covering the rotational excitations up to $J_{\rm max} = 15$  with the lower state energies ($\tilde{X}$) truncated at $hc \cdot 25000$~\cm\ and the upper state energies truncated at  $hc \cdot 28000$~\cm.

\begin{figure}[ht]
\centering
\includegraphics[width=0.7\columnwidth]{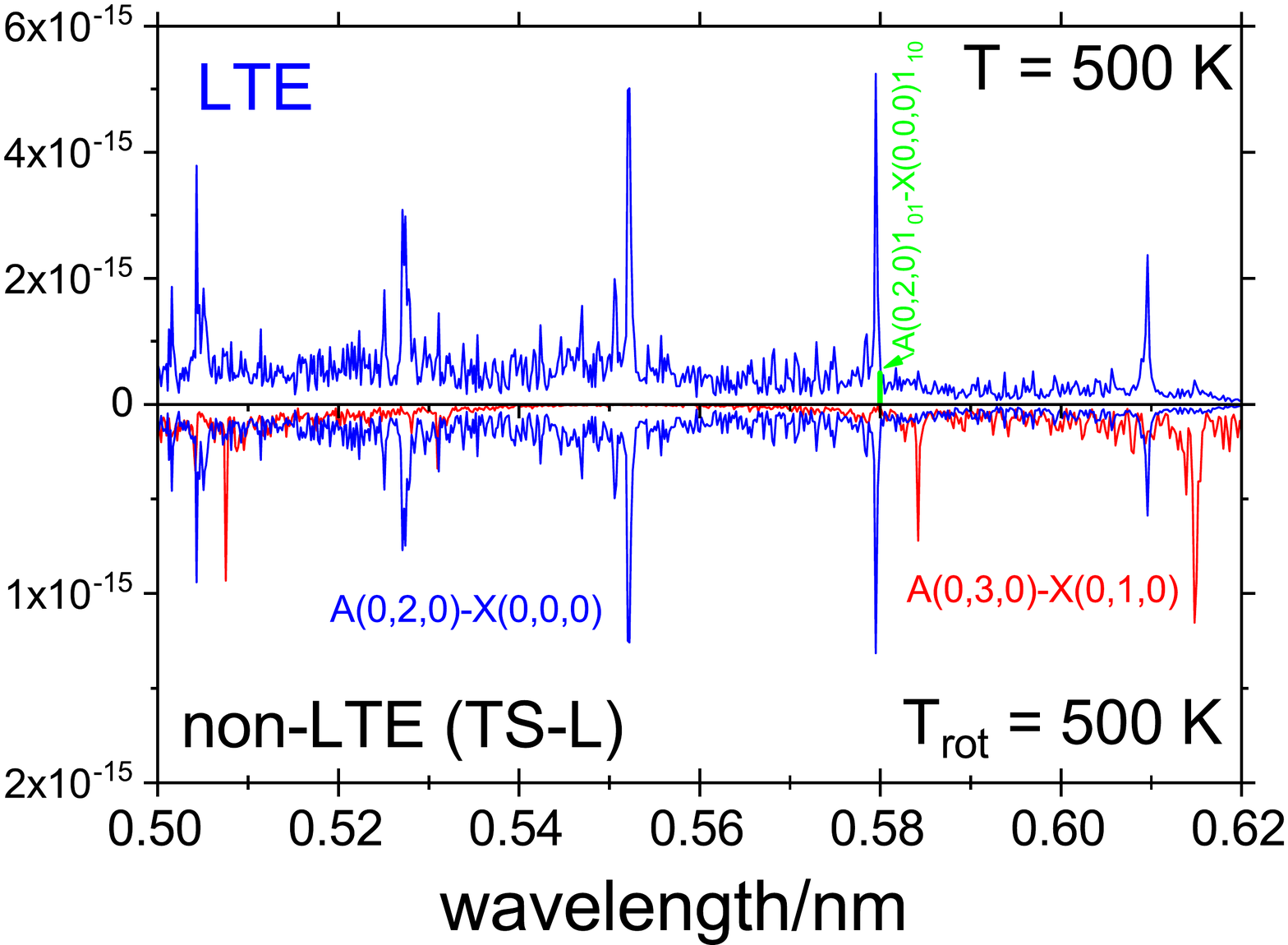}
\caption{An \ab\ -- \xa\ spectra of SiH\2, LTE ($T=500$~K) and non-LTE $T_{\rm rot} = 500$~K using the 1D model for Fragment TS-L. The energies and Einstein coefficients  are generated using the RENNER program with the spectroscopic model from \citet{04YuBuKr.SiH2}  A Gaussian line profile of HWHM=5 \cm. The popular ro-vibronic line $(0,0,0) 1_{01} \leftarrow (0,0,0) 1_{10}$ used in measurements of reaction rates of  SiH$_n$ species \citep{93KoKoOk.SiH2,95NoAkKo.SiH2,00HeJoxx.SiH2}  is shown }
\label{f:A-X}
\end{figure}

For the non-LTE simulations we used the 1D vibrational population model with the structural parameters corresponding to TS-L from Table~\ref{t:structures}.
Figure~\ref{f:A-X} shows a non-LTE electronic  spectrum of SiH\2\
in the region of the band  $\tilde{A}$ (0,2,0) $\leftarrow$ $\tilde{X}$ (0,0,0), assuming the rotational temperature $T_{\rm rot} = 500$~K, compared to an LTE spectrum of $T=$ 500~K. The non-LTE  spectrum contains the hot band $\tilde{A}$ (0,3,0) $\leftarrow$ $\tilde{X}$ (0,1,0) which can be used to identify the non-LTE character of the system. The rovibronic line $1_{01} \leftarrow 1_{10}$ belonging to this band   used in a number of experimental studies involving SiH\2\ as a reaction product  \citep{93KoKoOk.SiH2,95NoAkKo.SiH2,00HeJoxx.SiH2} to estimate reaction rates. It is common for such studies to assume the Boltzmann equilibrium  at different stages of the analysis of the measurements.   For example, the partition function of SiH\2\ is required to estimate the number density of SiH\2\ in its lower, ground electronic state \citep{94MaRoxx.SiH2,98CaRoSa.SiH2,00HeJoxx.SiH2}, which is directly affected by the LTE assumption. As our calculations show, the number densities of SiH\2\ as a reaction product can vary significantly under non-LTE depending on the reaction pathway and impact the experimental rates.

\section{Conclusion}
\label{sec:conc}

The focus of this paper is on the new features which have been added to \trove\ to allow modelling the non-LTE populations of polyatomic molecules.  We have demonstrated this capability by modelling the non-LTE line list of \sihh\ calculated with 3D wavefunctions and \trove, and compared them to non-LTE spectra of \sihh\ modelled using a 1D harmonic approach, and the LTE line list calculated previously by ExoMol.

There are two stable isomers of disilane, a local minimum structure and a global minimum structure, with a third transition state structure also known.  Non-LTE spectra of
SiH\2\ corresponding to dissociation  of disilane  from different sides of the three disilane isomer were computed. We have shown that the non-LTE spectra of \sihh\ are different in most cases.  This is important as the spectrum of \sihh\ is used to monitor the quantity present in a reaction as a means to track the progress of \forward\ and \reverse\ when calculating the corresponding rate constant. If  \disil\ is decomposing at a rate slower than it is being formed, then tracking the quantity of \sihh\ can give a rate constant that is not reflective of the speed of reaction, and merely an indication of the equilibrium balance of the two species \sihh\ and \disil.

In two approaches considered,  1D and the 3D, we assume that the rotational degrees of freedom are equillibrated quickly once the dissociation from disilane occurs, hence we use the Boltzmann distribution for the rotational degrees of freedom.  We also assume that the \sihh\ fragment during the instantaneous dissociation and is fully decoupled from the rest of the  \disil\ molecule, i.e. can be described by a 3D wavefunction in its lowest, relaxed vibrational configuration and has the same structural parameters as the \disil\ molecule.

We have shown that the non-LTE spectra of \sihh\ can be calculated by the new \trove\ methodology and existing ExoMol line list, and it compares well to the simpler 1D harmonic approximation published previously.  The method could be applied to the non-LTE spectroscopy of other small molecules including \silane, which has not been explored here.

We have also shown that despite the many approximations used in the 1D approximation (separability of the modes, Harmonic approximate etc.), the results compare well to the results obtained using the full 3D approach.  This lends  confidence in using the simplified but robust 1D approach in similar non-LTE studies, as e.g. we have used to model the CO non-LTE spectra,\citep{21PaCiCl.CO} recently, which are planning to explore in the future.

The methods described here can be used model the intensity distribution of the
reaction products and to ascertain from what molecule the \sihh\ dissociated from. The equilibrium structured parameters (bond lengths and angles) can be treated as effective parameters to be adjusted to reproduce the experimental spectra.

\section*{Acknowledgements}

This work was supported by UK research councils EPSRC, under grant EP/N509577/1 with COVID extension, and  STFC, under grant ST/R000476/1. This work made extensive use of the STFC DiRAC HPC facility supported by BIS National E-infrastructure capital grant ST/J005673/1 and STFC grants ST/H008586/1 and ST/K00333X/1. We thank the European Research Council (ERC) under the European Union’s Horizon 2020  research and innovation programme through Advance Grant number  883830. We also thank Thomas Mellor for help with the variational model.



\providecommand*{\mcitethebibliography}{\thebibliography}
\csname @ifundefined\endcsname{endmcitethebibliography}
{\let\endmcitethebibliography\endthebibliography}{}

\end{document}